\documentclass[10pt,twocolumn,letterpaper]{article}

\usepackage{cvpr}              
\usepackage{multirow}
\usepackage{pifont}
\usepackage[accsupp]{axessibility}
\newcommand{\cmark}{\ding{51}}%
\newcommand{\xmark}{\ding{55}}%

\definecolor{cvprblue}{rgb}{0.21,0.49,0.74}
\usepackage[pagebackref,breaklinks,colorlinks,allcolors=cvprblue]{hyperref}

\newcommand{\myparagraph}[1]{\smallskip\noindent\textbf{#1}}

\def\paperID{6820} 
\def\confName{CVPR}
\def\confYear{2025}

\title{
\emph{TopoCellGen}: Generating Histopathology Cell Topology with a Diffusion Model
}

\author{
Meilong Xu$^{1}$ \quad
Saumya Gupta$^{1}$ \quad
Xiaoling Hu$^{2}$ \quad
Chen Li$^{1}$ \quad
Shahira Abousamra$^{3}$ \\
Dimitris Samaras$^{1}$ \quad
Prateek Prasanna$^{1}$ \quad
Chao Chen$^{1}$ \\[3pt]
$^{1}$Stony Brook University, NY, USA \\
$^{2}$Athinoula A. Martinos Center for Biomedical Imaging, \\
Massachusetts General Hospital and Harvard Medical School, MA, USA \\
$^{3}$Department of Biomedical Data Science, Stanford University, CA, USA \\
{\tt\small meixu@cs.stonybrook.edu, chao.chen.1@stonybrook.edu}
}

\begin{document}
\maketitle

\begin{abstract}
Accurately modeling multi-class cell topology is crucial in digital pathology, as it provides critical insights into tissue structure and pathology. The synthetic generation of cell topology enables realistic simulations of complex tissue environments, enhances downstream tasks by augmenting training data, aligns more closely with pathologists' domain knowledge, and offers new opportunities for controlling and generalizing the tumor microenvironment.
In this paper, we propose a novel approach that integrates topological constraints into a diffusion model to improve the generation of realistic, contextually accurate cell topologies. Our method refines the simulation of cell distributions and interactions, increasing the precision and interpretability of results in downstream tasks such as cell detection and classification.
To assess the topological fidelity of generated layouts, we introduce a new metric, Topological Fréchet Distance (TopoFD), which overcomes the limitations of traditional metrics like FID in evaluating topological structure.
Experimental results demonstrate the effectiveness of our approach in generating multi-class cell layouts that capture intricate topological relationships. Code is available at \href{https://github.com/Melon-Xu/TopoCellGen}{https://github.com/Melon-Xu/TopoCellGen}.
\end{abstract}

\section{Introduction}
Deep-learning methods have made substantial advances in foundational tasks for nuclei analysis, including instance segmentation~\cite{he2023toposeg, graham2019hover, koohbanani2020nuclick, horst2024cellvit, you2024mine}, classification, and detection~\cite{abousamra2021multi, prangemeier2020attention}. These tasks provide the basis for downstream analyses, enabling detailed characterization of tissue architecture and cellular interactions, which are crucial for diagnostic and prognostic applications in pathology~\cite{lu2021feature, ding2022image}. However, accurately annotating multi-class cell arrangements in pathology images is challenging, as distinct cell types display unique spatial patterns, requiring significant domain expertise. Although annotated datasets for multiple cell types exist, they often lack the diversity needed for generalization across various tissues and organs. 

To alleviate the burden of manual annotation and enhance the efficiency of analysis, there has been growing interest in utilizing generative models. Early works use Generative Adversarial Networks (GANs)~\cite{goodfellow2020generative} for automatic generation of pathology images~\cite{abousamra2023topology,hou2019robust,deshpande2024synclay}. 
In recent years, diffusion models~\cite{sohl2015deep, ho2020denoising, nichol2021improved, dhariwal2021diffusion, zhang2023adding, zhao2024uni, li2025controlnet, wang2023inversesr} have emerged as much more reliable alternatives, 
generating accurate, high-resolution histopathology images~\cite{yellapragada2024pathldm, oh2023diffmix, min2024co, graikos2024learned, aversa2024diffinfinite}. 
However, all these diffusion models are only trained to directly generate the histopathology images. 
Despite the impressive visual results, these powerful models deliver limited insight into the underlying biology. It is very hard to connect the learned distributions with human knowledge about tumor microenvionment. This makes it challenging to validate, generalize, or control these models. 

We argue that a key issue is the lack of explicit generation of cells and their spatial arrangement. The spatial organization of cells and the interactions across different cell types are critical for understanding tumor microenvironments, disease progression, and tissue regeneration~\cite{schapiro2017histocat}. The density and spatial distribution of various cell types -such as lymphocytes, epithelial cells, and stromal cells- are essential for pathologists in making accurate diagnoses and prognoses. For instance, the detection and quantification of tumor-infiltrating lymphocytes (TILs), which are lymphocytes located within the boundaries of invasive tumors~\cite{salgado2015evaluation}, have been strongly linked to improved clinical outcomes~\cite{stanton2016clinical, saltz2018spatial}. The presence of isolated or small clusters of tumor cells at the invasive front, known as tumor budding, serves as a prognostic biomarker associated with a higher risk of lymph node metastasis in colorectal carcinoma and other solid malignancies~\cite{lugli2021tumour}.

In this paper, we explore the problem of generating cell spatial layouts, with multiple benefits. First and foremost, directly modeling cell layouts brings us closer to aligning with pathologists' domain knowledge. This enables the direct verification of synthetic data through quantification and comparison with expert knowledge, facilitating greater accuracy and trust in the generated data. Additionally, it opens up the possibility of controlling the layout generation process, allowing the model to generalize to previously unseen scenarios. From a data augmentation perspective, generating cell layouts also enables the creation of histology images conditioned on these layouts. This capability allows for the production of synthetic images with cell annotations, which can significantly aid in the training of models for various downstream tasks, particularly cell detection and classification.

\begin{figure}[tbp]
    \centering
    \includegraphics[width=0.49\textwidth]{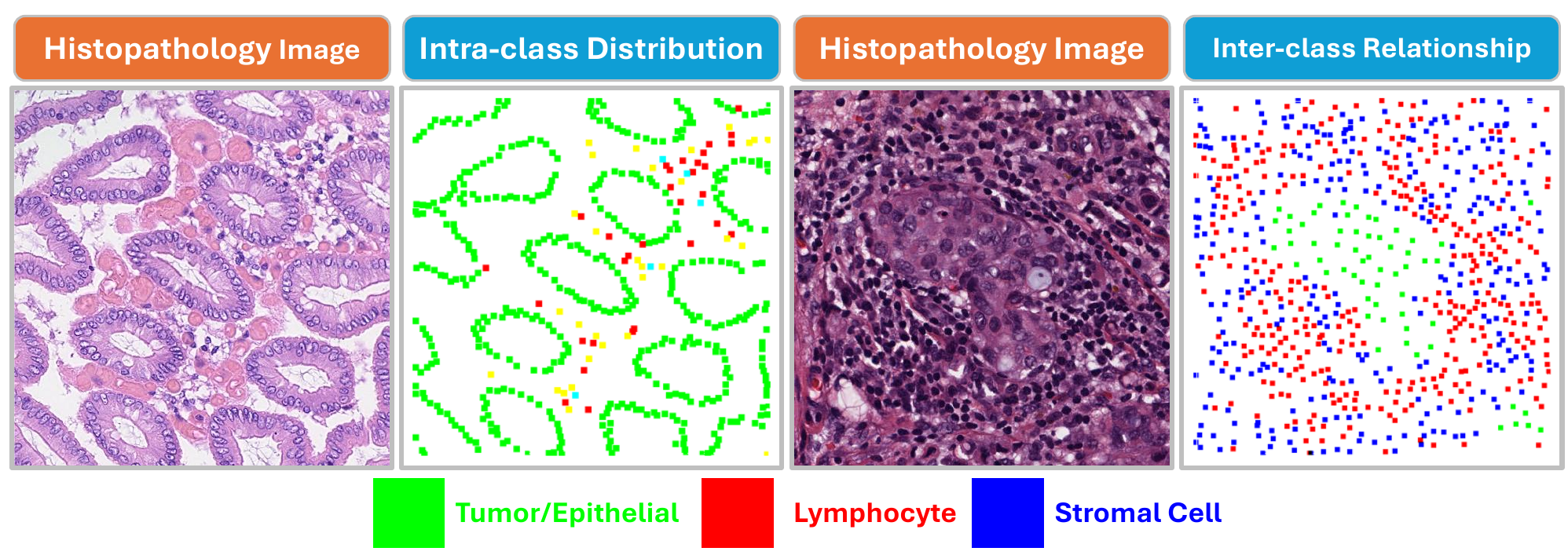}
    \vspace{-.2in}
    \caption{Illustrations of intra-class distribution and the inter-class relationship across various cell types. Here we only highlight the \textcolor{green}{tumor/epithelial}, \textcolor{red}{lymphocytes}, and \textcolor{blue}{stromal cells}.
    }
    \vspace{-.2in}
    \label{fig:motivation}
\end{figure}

\myparagraph{Cell topology is the key.}
To generate accurate cell layouts, we argue that the key is to model the topological relationships between cells, defined by their spatial organization, clustering, mixing, and connectivity. These topological patterns provide valuable insights into cellular communication, structural changes, and morphological abnormalities linked to pathological conditions such as cancer, inflammation, and fibrosis.
\cref{fig:motivation} illustrates the significance of these topological relationships in understanding tissue organization and disease progression. These relationships reveal how cells are spatially arranged to form structural motifs that are crucial for tissue function and stability. For example, on the left, we show that in healthy glandular tissues, epithelial cells typically organize into circular or tubular structures, forming acinar units that are essential for secretion and nutrient transport~\cite{schapiro2017histocat}.
In tissues with diverse cell types, topological relationships capture spatial configurations and connectivity patterns that reveal cellular interactions critical for tissue health or disease~\cite{yuan2016spatial}. For example, immune cells clustering around tumor cells (\cref{fig:motivation} (right)) or specific fibroblast-epithelial arrangements highlight processes like immune surveillance, inflammation, and stromal support. By identifying these patterns, topological methods offer insights into multi-class cellular interactions, revealing potential biomarkers and enhancing our understanding of disease progression.

\myparagraph{We propose the first diffusion model that generates cell topology for digital pathology.}
Our method guides the generation process with both 0- and 1-dimensional topological features, i.e., cell clusters and holes/gaps enclosed by cells. We implement this using the theory of persistent homology, which models cell topology in a multi-scale manner. As illustrated in \cref{fig:motivation}, learning the topology ensures that the generated cell layouts not only preserve intra-class spatial/structural property but also accurately capture inter-class interactions. 

The second contribution of this paper is a novel cell counting loss that ensures the diffusion model learns the correct distribution of cell numbers from the data. This addresses a key challenge in previous diffusion models for histopathology images~\cite{graikos2024learned, aversa2024diffinfinite, li2024spatial}, which is often biased toward unrealistically low or high cell counts. 

By ensuring both accurate cell topology and cell count, our method generates layouts that more closely resemble real tissue microenvironments. This precision in modeling cell distributions and interactions enhances the utility of generated images for augmenting training in downstream tasks such as cell detection and classification, as demonstrated in our experiments. From a modeling perspective, our approach moves closer to generating biologically faithful representations of cellular environments, integrating human domain knowledge, and improving generalization to diverse scenarios.

Finally, we introduce the Topological Fréchet Distance (TopoFD), a novel metric designed to assess the spatial and topological accuracy of the generated layouts. TopoFD measures the similarity between the topological features of real and synthetic cell layouts, whereas traditional metrics such as FID focus solely on visual similarity. 

In summary, our contribution is three-fold:
\begin{itemize}
    \item We present the first diffusion model designed to generate cell topology for digital pathology, simulating realistic intra- and inter-class spatial distributions in the generated layouts. 
    \item We introduce a novel cell counting loss that aligns the generated cell numbers with real data, addressing biases in prior diffusion models and ensuring realistic cell density in synthetic layouts.
    \item We introduce the Topological Fréchet Distance (TopoFD), a new metric designed to evaluate the topological similarity of generated cell layouts.
\end{itemize}
Extensive experiments demonstrate the effectiveness of our proposed method, showing that it not only enhances sample quality but also significantly improves performance in downstream tasks such as cell detection and classification.

\section{Related Works}
\myparagraph{Diffusion Models for Digital Pathology.}
Diffusion models, such as Denoising Diffusion Probabilistic Models (DDPMs)~\cite{ho2020denoising, sohl2015deep, nichol2021improved, song2020denoising, dhariwal2021diffusion, zhao2023ddfm, fei2023generative, hou2024global} and Latent Diffusion Models (LDMs)~\cite{rombach2022high, saharia2022photorealistic, brooks2023instructpix2pix}, have significantly advanced image synthesis by modeling data distributions through iterative denoising processes on data sample or latent spaces. These models have been adapted to generate high-fidelity histopathology images in digital pathology~\cite{yellapragada2024pathldm, aversa2024diffinfinite, graikos2024learned, harb2024diffusion}. 
However, these methods often overlook the significance of multi-class cell layouts, which are crucial for accurately representing tissue structures.

Recent approaches to synthetic cell layout generation in digital pathology have explored nuclei labeling and cell arrangement. Abousamra et al. utilized spatial statistics and topological descriptors in GANs to model complex cell configurations~\cite{abousamra2023topology}. DiffMix employs conditioned diffusion models for augmenting imbalanced nuclear pathology datasets, generating more realistic images than earlier methods~\cite{oh2023diffmix}. Similarly, a inter diffusion framework proposed in~\cite{min2024co} generates paired histopathology images and nuclei labels simultaneously, enhancing the context-awareness of synthetic data. Another strategy, a diffusion model focused on cell layout generation, uses density maps to incorporate spatial distributions, though it simplifies cell counts into five categorical conditions, which provides only limited guidance~\cite{li2024spatial}.
While these models have contributed to more realistic cell layout synthesis, they generally lack fine control over cell density and fail to explicitly preserve intra-class spatial distributions and inter-class spatial relationships across cell types. This limits their effectiveness in replicating the complex and varied structures of real tissue.

\myparagraph{Topology-Driven Methods in Deep Learning.}
Incorporating algebraic topology~\cite{munkres1984elements} in deep learning frameworks is becoming increasingly significant.
A key development in this front has been persistent homology~\cite{edelsbrunner2002topological, edelsbrunner2022computational}, a mathematical theory that analyzes how structural features (such as connected components, loops, and voids) persist across different scales in data. This approach has proven particularly valuable for understanding complex data structures while being resistant to noise. Its applications range from image segmentation~\cite{hu2019topology, clough2020topological, xu2025semi, stucki2023topologically}, feature extraction~\cite{gabrielsson2020topology} and disease diagnosis~\cite{wang2024topotxr, levenson2024advancing, xu2024topology, lee2008implications}.  
The field has further expanded through additional mathematical frameworks, including discrete Morse theory~\cite{forman1998morse, hu2020topology, hu2022learning}, homotopy warping~\cite{hu2022structure}, structural relationship studies~\cite{gupta2022learning, gupta2024topology}, and methods for analyzing shape features~\cite{shit2021cldice, wang2022ta}. 

For generative models, topological information has become increasingly essential, particularly in biomedical imaging, where capturing both visual realism and topological integrity is critical.
One notable method is TopoGAN~\cite{wang2020topogan}, which integrates topology into GANs by introducing a novel topological loss based on persistent homology. Recently, TopoDiffusionNet~\cite{gupta2024topodiffusionnet} has integrated topology with diffusion models, introducing a topology-based objective function that guides the model's denoising process to generate images with exact object counts. 
However, these methods, designed for natural images, are unsuitable for our cell topology generation task. Their emphasis on overall structure numbers cannot capture the cell structural pattern and cell spatial interactions among multiple cell types as our method does.

\begin{figure*}[t]
\centering
    \includegraphics[width=\linewidth]{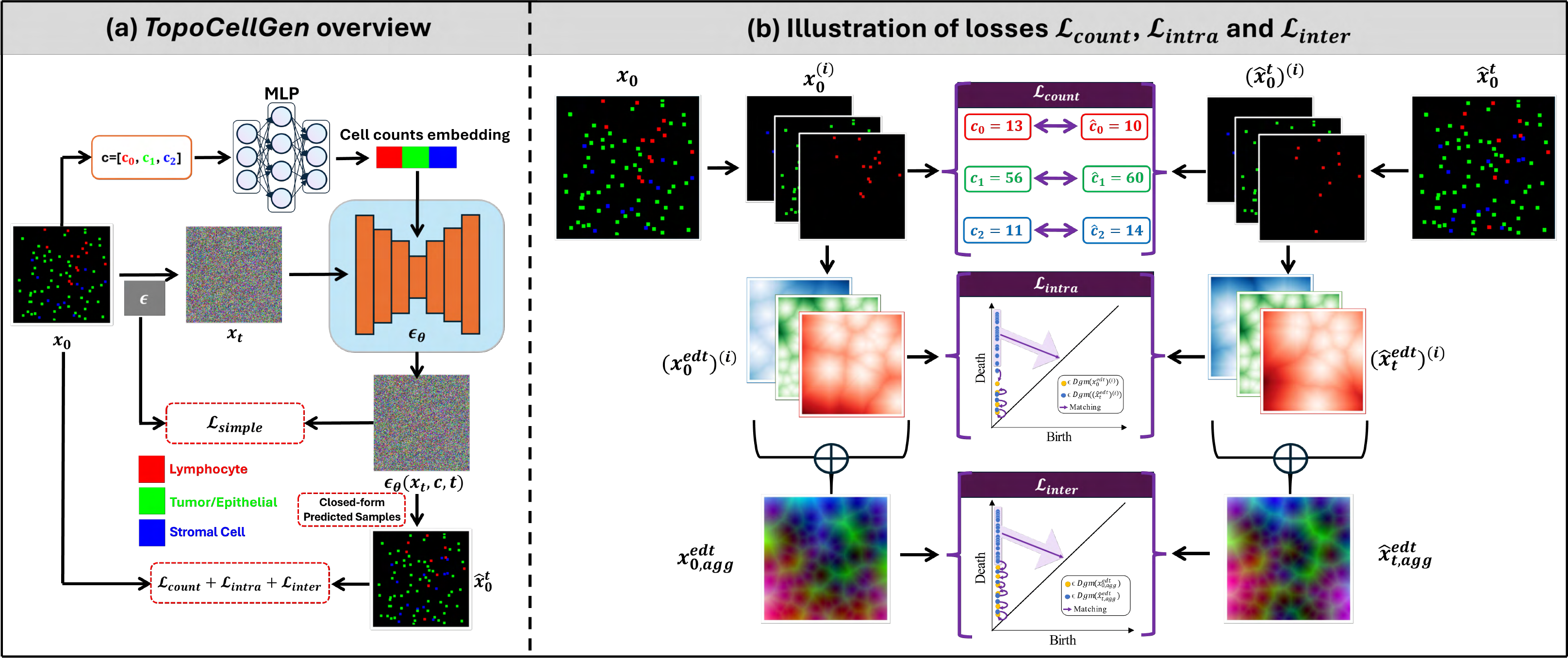}
    \vspace{-.2in}
    \caption{An overview of our method \emph{TopoCellGen}. (\textbf{a}) denotes the overview workflow. (\textbf{b}) shows the details of $\mathcal{L}_{\text{count}}$, $\mathcal{L}_{\text{intra}}$ and $\mathcal{L}_{\text{inter}}$.
    }
    \vspace{-.1in}
    \label{fig:merged-overview}
\end{figure*}

\section{Method}
In this section, we introduce our methodology for generating realistic multi-class cell layouts with precise topological control. To accurately control the cell density in the generated layout, we employ a conditioning mechanism based on the cell count for each type. In the multi-class setting, assuming we have $n$ classes of cells, we define the condition vector $c=[c_1, c_2, ..., c_n]$, where each element $c_i$ represents the count of cells in the respective channel.

However, the condition vector $c$ alone is insufficient to precisely control cell counts~\cite{gupta2024topodiffusionnet}. Additionally, it cannot preserve the intra-class spatial distributions as well as inter-class topological relationships in the generated layouts. To address this, we formulate a cell counting loss to strengthen the control on accurate cell counts and introduce two topology-aware objective functions based on persistent homology~\cite{edelsbrunner2002topological, munkres1984elements} to guide the generation process, ensuring realistic and contextually accurate cell layouts that respect both spatial and topological constraints.

The remainder of this section is structured as follows: In~\cref{subsec:diffusion_models}, we present a brief introduction to diffusion models. Next, in~\cref{subsec:persistent_homology}, we provide a quick background on persistent homology. Following this, we introduce \emph{TopoCellGen}, our topology-preserving cell layout generation method, in~\cref{subsec:TopoCellGen}. We introduce our proposed Topological Fréchet Distance (TopoFD) in~\cref{subsec:TopoFD}. Finally, in~\cref{subsec:layout-to-image}, we present the overall inference pipeline and how the generated cell layouts are transformed into H\&E images for training augmentation for downstream tasks.

\myparagraph{Preliminaries. }In a multi-class cell layout containing $n$ channels, each channel corresponds to a specific cell type (e.g., lymphocyte), with each cell represented as a square where the pixel value is set to 1, while the pixel value of the background is set to $0$. 

\subsection{Diffusion Models}
\label{subsec:diffusion_models}
Our generative approach utilizes a denoising diffusion probabilistic model (DDPM)~\cite{ho2020denoising}, which learns to reverse a forward process that incrementally adds Gaussian noise to transform a structured cell layout into a noise distribution. The reverse process then reconstructs the layout from noise via iterative denoising.

Let $x_0$ represent the target cell layout and $x_T$ denote pure Gaussian noise. 
At each time step $t$, noise is added to the data based on a variance schedule $\beta_t$: $q(x_t | x_{t-1}) = \mathcal{N}(x_t; \sqrt{1-\beta_t} x_{t-1}, \beta_t I)$. This forward process results in the progressively noisier version of the data, with $x_T$ approximating an isotropic Gaussian distribution as $t$ increases.

The reverse process, parameterized by a neural network $\epsilon_{\theta}(x_t, t)$ (typically UNet~\cite{ronneberger2015u}), learns to iteratively denoise $x_T$ back to $x_0$. Conditioning on the cell count vector $c$, the model is trained by minimizing a simplified variant of the variational lower bound, specifically focusing on predicting the noise added at each step:
\begin{equation}
    \mathcal{L}_{\text{simple}} = \mathbb{E}_{t, x_0, \epsilon} \left[ \| \epsilon - \epsilon_\theta(x_t, c, t) \|^2 \right]
\end{equation}
where $\epsilon \sim \mathcal{N}(0, I)$ is the noise sampled during training. This objective enables the model to learn the reverse process effectively. Instead of the standard iterative denoising, we also approximate the noiseless layout $\widehat{x}_0^t$ deterministically for any noisy state $x_t$ by marginalizing over the noise schedule: 
\vspace{-.1in}
\begin{equation}
    \widehat{x}_0^t \approx \frac{1}{\sqrt{\bar{\alpha}_t}} \left( x_t - \sqrt{1 - \bar{\alpha}_t} \epsilon_\theta(x_t, c, t) \right)
    \vspace{-.1in}
\label{equl:predicted_noiseless_layout}
\end{equation}
where $\alpha_t=1-\beta_t$ and $\bar{\alpha}_t = \prod_{s=1}^t \alpha_s$, which aggregates the effect of the variance schedule up to time $t$.
This predicted noiseless layout, $\widehat{x}_0^t$, will be used to impose constraints in subsequent stages.

\subsection{Background: Persistent Homology}
\label{subsec:persistent_homology}
In algebraic topology~\cite{munkres1984elements}, homology classes provide a structured way to capture topological features across multiple dimensions. For instance, 0-, 1-, and 2-dimensional features represent connected components, loops, and voids, respectively. The $d$-dimensional Betti number, $\beta_d$, quantifies the number of $d$-dimensional features present, offering insights into the underlying topological complexity. However, extending these concepts to real-world data, which is often continuous and noisy, introduces challenges for accurately capturing topological structures.

Persistent homology, developed in the early 2000s~\cite{edelsbrunner2002topological, edelsbrunner2022computational}, addresses this need by tracking the evolution of topological features across multiple scales, making it particularly effective for discrete datasets like cell point clouds. The process begins by constructing a filtration—a sequence of nested simplicial complexes built from the point cloud data by incrementally connecting points based on a scale parameter. As the parameter varies, topological features such as connected components and loops appear and eventually vanish, each represented by a point in a persistence diagram (Dgm). In a diagram, each point $(b, d)$ marks the birth and death of a feature, capturing its persistence across scales and providing a compact, multi-scale summary of the underlying topology. More details are in the Supplementary.

\subsection{Spatially Aligned Cell Layout Generation}
\label{subsec:TopoCellGen}
The primary objective of our method is to generate multi-class cell layouts that accurately simulate both the topological and spatial properties of real-world biological cell distributions. To achieve this, we ensure accurate cell counts for each cell type through a cell counting loss, while also preserving spatial relationships within individual cell types via enforcing intra-class spatial consistency. Furthermore, we maintain structural coherence across all cell types by applying an inter-class structural regularization, leveraging 1-dimensional persistent homology to encapsulate both type-specific and collective spatial properties. The overall pipeline is shown in \cref{fig:merged-overview}.

\myparagraph{Cell Counting Loss.} Given the target layout $x_0$, which serves as the ground truth, for each time step $t$, we obtain the predicted noiseless layout $\widehat{x}_0^t$ using~\cref{equl:predicted_noiseless_layout}. To ensure precise control over the number of cells in the generated layout, we introduce a differentiable cell counting loss. The key challenge lies in making the counting operation differentiable for gradient-based optimization. We address this by employing the Straight-Through Estimator (STE)~\cite{bengio2013estimating}, which enables gradient flow through the discrete binarization operation. Specifically, after obtaining $\widehat{x}_0^t$, 
we apply a hard threshold to obtain binary values:
\begin{equation}
b(\widehat{x}_0^t) = float((\widehat{x}_0^t \geq \tau))
\end{equation}
where $\tau$ is the threshold parameter. Here we set it to the median value of $\widehat{x}_0^t$. During back-propagation, the STE treats the thresholding operation as an identity function, allowing gradients to flow through.
The cell counting loss is then formulated as:
\vspace{-.1in}
\begin{equation}
\mathcal{L}_{\text{count}} = \frac{1}{|n|}\sum_{i=1}^{n} \left|\frac{\sum b(\widehat{x}_0^t)^{(i)}}{\delta} - \frac{\sum x_0^{(i)}}{\delta}\right|
\vspace{-.1in}
\end{equation}
where $b(\widehat{x}_0^t)^{(i)}$ represents the binarized prediction for the $i$-th channel, and $\delta$ indicates the area ($3\times 3$) of a single cell in the layouts. This formulation provides a differentiable approximation to the discrete cell counting operation, enabling end-to-end training while maintaining precise control over the number of cells for each cell type. 

\myparagraph{Intra-Class Spatial Consistency.}
To enforce spatial consistency within each cell type, we first calculate the distance transform map~\cite{felzenszwalb2012distance} for each channel in both the target layout $x_0$ and the predicted noiseless layout $\widehat{x}_0^t$. The distance transform $D(x)$ is a function that assigns to each pixel the minimum Euclidean distance to the nearest cell (or non-zero pixel) in the channel. This can be formally written as:
\begin{equation}
    D(x) = \min_{p \in \text{cells}} \| x - p \|
    \vspace{-.1in}
\end{equation}
where $p$ represents the positions of cells in the layout. After obtaining the distance transform maps of the target layout and the predicted noiseless layout, $\widehat{x}^{edt}_t=D(b(\widehat{x}_0^t))$ and $x^{edt}_0=D(x_0)$, we calculate the 1-dim persistence diagrams for both of them, $Dgm(\widehat{x}^{edt}_t)$ and $Dgm(x^{edt}_0)$ respectively. Similar to previous topological losses \cite{hu2019topology}, we will use the classic Wasserstein distance between the two diagrams. Given two diagrams $Dgm(q)$ and $Dgm(s)$, the $p$-th Wasserstein distance is defined as follows:
\vspace{-.1in}
\begin{equation}
    W_p(Dgm(q), Dgm(s))=\left( \underset{\gamma\in \Gamma}{inf}\sum\limits_{x \in Dgm(q)}{||x-\gamma(x)||}^p \right)^{\frac{1}{p}}
    \nonumber
    \vspace{-.1in}
\end{equation}
where $\Gamma$ represents all bijections from $Dgm(q)$ to $Dgm(s)$.

The Wasserstein distance operates by identifying an optimal correspondence between points in two diagrams, assigning unmatched points to their projections on the diagonal. This distance metric is calculated by summing the distances between all paired points. The process of finding this optimal matching, as well as calculating the Wasserstein distance, can be accomplished using either the traditional Hungarian algorithm or more sophisticated methods~\cite{kerber2016geometry, lacombe2018large}.

Next, we denote $\gamma^\ast$, the optimal matching between $Dgm(\widehat{x}^{edt}_t)$ and $Dgm(x^{edt}_0)$. Each persistence dot in $Dgm(\widehat{x}^{edt}_t)$ is matched either to a target dot in $Dgm(x^{edt}_0)$ or its projection on the diagonal. We can now formulate the spatial distribution consistency loss as the squared distance between every dot in $Dgm(\widehat{x}^{edt}_t)$ and its match:
\vspace{-.1in}
\begin{equation}
     \mathcal{L}_{\text{spc}} = \sum\limits_{q \in Dgm(\widehat{x}^{edt}_t)}||q-\gamma^\ast(q)||^2
     \vspace{-.1in}
     \label{eq:spa-cons}
\end{equation}

For a multi-class cell layout containing $n$ classes of cells, we formulate the intra-class spatial consistency loss as follows by averaging the \cref{eq:spa-cons} across multiple classes:
\vspace{-.1in}
\begin{equation}
    \mathcal{L}_{\text{intra}} = \frac{1}{|n|} \sum_{i=1}^{n} \mathcal{L}_{\text{spc}} \left( Dgm((\widehat{x}^{edt}_t)^{(i)})), Dgm((x^{edt}_0)^{(i)}) \right)
\end{equation}

\myparagraph{Inter-Class Structural Regularization.}
Beyond maintaining spatial distribution consistency within individual cell types, it is equally important to capture the relationships between different cell types. To achieve this, we construct a unified layout by combining all cell types into a single-channel representation, referred to as the aggregated layout:  $x_0^{agg}=Agg(x_0)$ and $\widehat{x}_{0}^{t,agg}=Agg(\widehat{x}_{0}^{t})$. We then compute the distance transform for the aggregated layouts, with $\widehat{x}^{edt}_{t,agg}=D(\widehat{x}_{0}^{t,agg})$ representing the distance transform of the predicted layout and $x^{edt}_{0,agg} = D(x_0^{agg})$ for the target layout.
The inter-class structural loss $\mathcal{L}_{\text{inter}}$ is computed similarly to the intra-class loss:
\begin{equation}
    \mathcal{L}_{\text{inter}} = \mathcal{L}_{\text{spc}} \left( 
    Dgm(\widehat{x}^{edt}_{t,agg}), Dgm(x^{edt}_{0,agg})
    \right)
    \vspace{-.1in}
\end{equation}
\noindent 
Together, these class-specific and cross-class regularizations ensure that both individual cell distributions and their cumulative spatial interactions are enforced, preserving critical spatial dynamics within and between cell types in the generated layouts.

\myparagraph{Final Objectives.}
The final training objective function of the model is the weighted sum of the three losses with $\mathcal{L}_{\text{simple}}$:
\vspace{-.1in}
\begin{equation}
    \mathcal{L}_{\text{total}} = \mathcal{L}_{\text{simple}} + \lambda_{c}\mathcal{L}_{\text{count}} + \lambda_{\text{intra}}\mathcal{L}_{\text{intra}} + \lambda_{\text{inter}}\mathcal{L}_{\text{inter}}
\end{equation}
where $\lambda_{c}$, $\lambda_{\text{intra}}$ and $\lambda_{\text{inter}}$ are hyper-parameters that control the relative contributions of the respective loss terms.

\subsection{Topological Fréchet Distance (TopoFD)}
\label{subsec:TopoFD}
Fréchet Inception Distance (FID)~\cite{heusel2017gans} is a key metric for evaluating the quality of generated image data by comparing the distributions of real and generated samples. FID operates by computing the mean and covariance of feature representations extracted from a pre-trained model, typically InceptionV3~\cite{szegedy2016rethinking}. The distance between these distributions is then measured using the Fréchet distance~\cite{frechet1957distance}. 
The formula for FID is as follows:
\begin{equation}
    \text{FID} = \|\mu_r - \mu_g\|^2 + \text{Tr}(\Sigma_r + \Sigma_g - 2(\Sigma_r \Sigma_g)^{\frac{1}{2}})
\end{equation}
where $\mu_r$, $\Sigma_r$ and $\mu_g$, $\Sigma_g$ are the mean and covariance of the real and generated data, respectively.

\begin{figure}[t]
    \centering
    \includegraphics[width=0.48\textwidth]{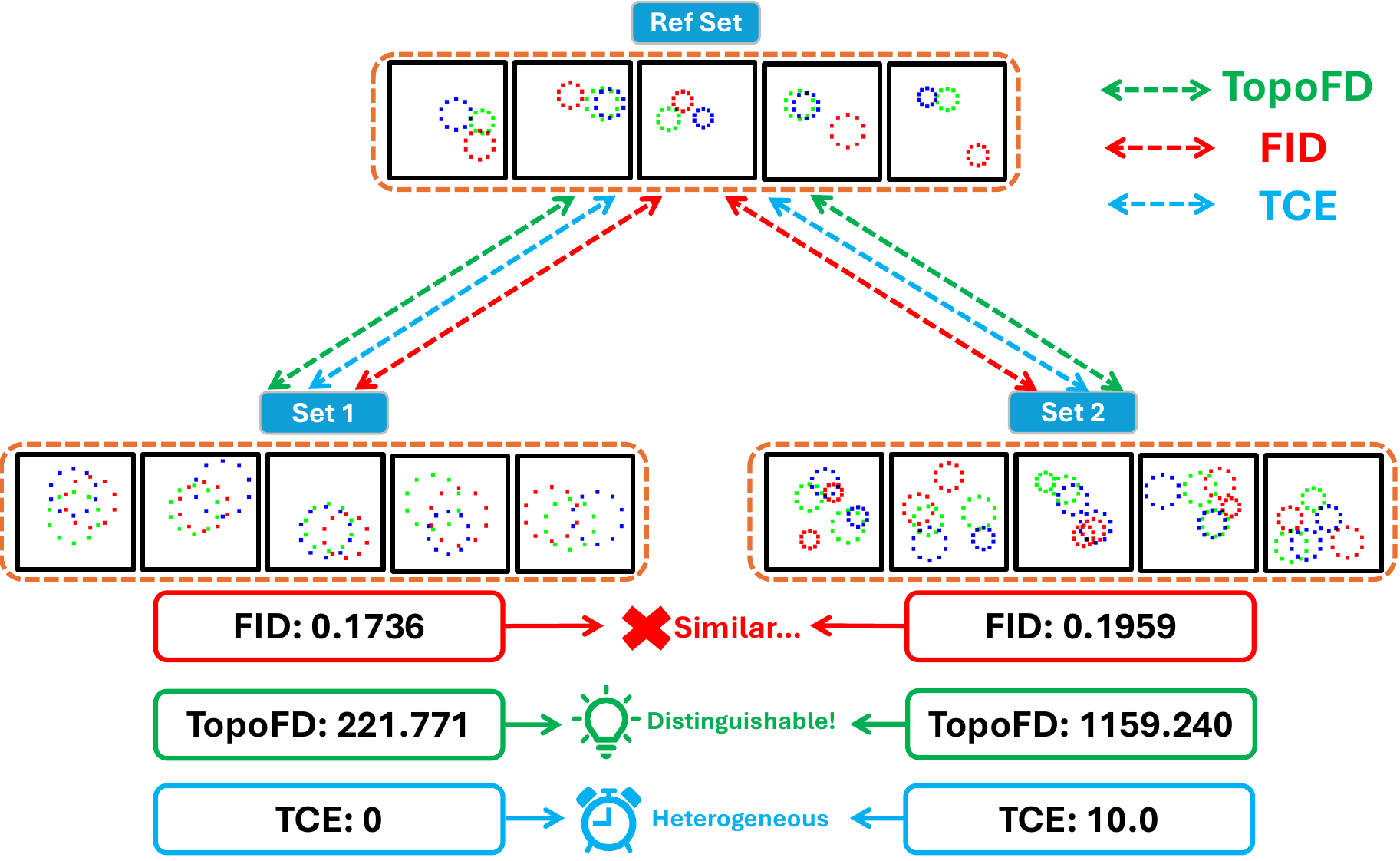}
    \vspace{-.1in}
    \caption{Intuition of our proposed Topological Fréchet Distance. TCE indicates the Total Count Error.}
    \vspace{-.1in}
    \label{fig:TopoFD_intuition_v2}
\end{figure}
Traditional FID measures feature similarities but fail to capture complex spatial and topological cell interactions. For example, in \cref{fig:TopoFD_intuition_v2}, synthetic layouts of three cell types (\textcolor{red}{lymphocyte}, \textcolor{green}{tumor/epithelial} and \textcolor{blue}{stromal cells}) 
demonstrate this limitation. Both Set 1 and Set 2 yield similar FID scores with respect to the Ref Set, yet differ significantly from each other in spatial configuration. Set 1 mirrors the reference with cohesive single cycles per cell type, while Set 2 presents multiple distinct cycles, deviating from the reference. The Total Count Error (TCE) reveals identical counts between Set 1 and the reference (TCE = 0) but substantial discrepancies for Set 2 (TCE = 10). Despite matching counts, Set 1's spatial arrangement still diverges from the reference. This suggests that a count metric alone does not ensure spatial and topological fidelity. We propose Topological Fréchet Distance (TopoFD) to capture higher-dimensional topological features, enabling greater sensitivity to spatial configurations essential in synthetic pathology data generation.

\begin{figure}[t]
    \centering
    \includegraphics[width=0.48\textwidth]{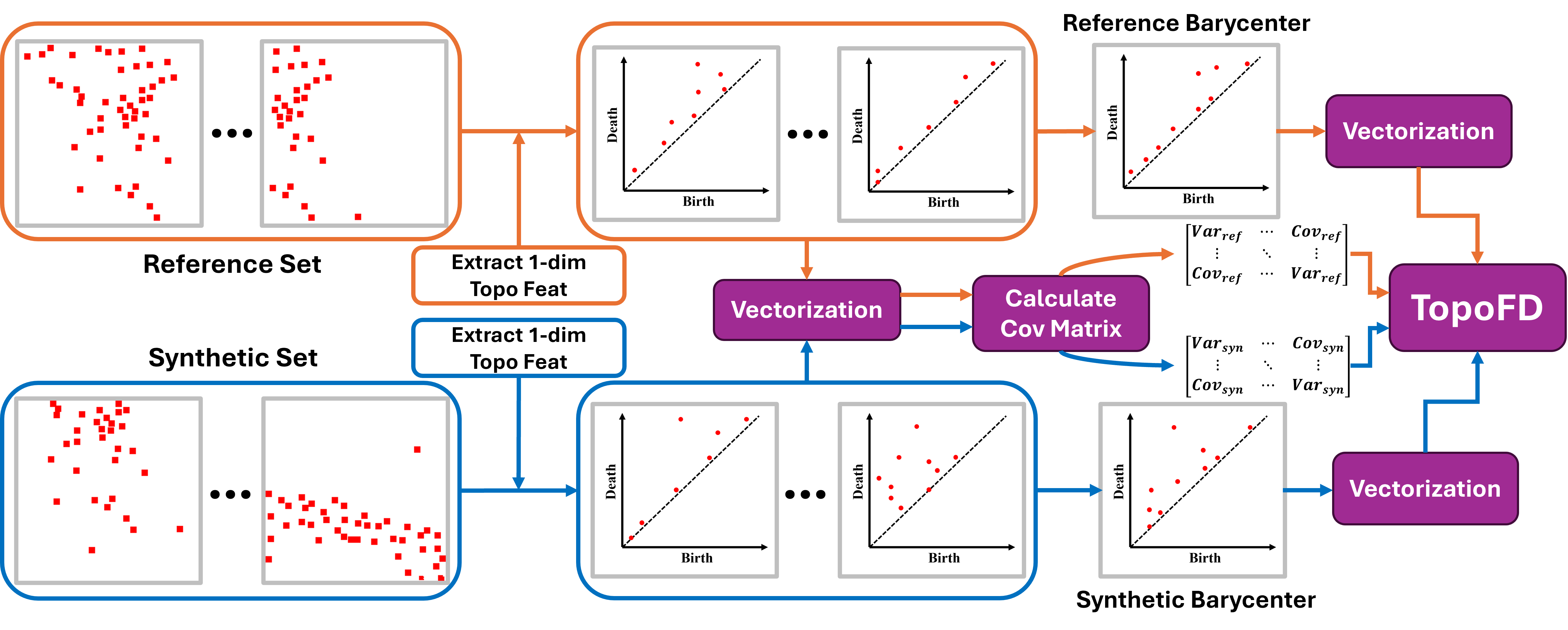}
    \vspace{-.1in}
    \caption{The overall pipeline of calculating the Topological Fréchet Distance. Take the \textcolor{red}{lymphocyte} as an example.}
    \vspace{-.1in}
    \label{fig:TopoFD_pipeline}
\end{figure}

The pipeline of our Topological Fréchet Distance (TopoFD) is shown in \cref{fig:TopoFD_pipeline}, taking one type of cell as an example. For each layout in the reference and synthetic sets, we first obtain a point cloud, with the points corresponding to the cell center coordinates. Then, we calculate 1-dimensional persistence diagrams of each point cloud in each set:
\begin{align*}
    Dgm_{ref} = \{ Dgm_r^1, Dgm_r^2, ..., Dgm_r^n\} \\ 
    Dgm_{syn} = \{ Dgm_s^1, Dgm_s^2, ..., Dgm_s^n\}
\end{align*}
where $Dgm_r^i$ and $Dgm_s^j$ are persistence diagrams in reference and synthetic sets, $n$ is the number of the samples. We then compute each set's barycenter~\cite{cohen2010lipschitz, bubenik2015statistical}, which minimizes the sum of Wasserstein distances~\cite{cohen2010lipschitz} between individual diagrams and the barycenter. For the reference set, the barycenter $\overline{Dgm}_{real}$ is given by:
\begin{equation}
    \overline{Dgm}_{r} = \arg\min_{\overline{Dgm}_{r}} \sum_{i=1}^{n} W_2^2(Dgm_{r}^i, \overline{Dgm}_{r})
\end{equation}
where $W_2$ is the 2-Wasserstein distance. A similar process is applied for $\overline{Dgm}_{s}$ from the synthetic set.

\begin{figure*}[t]
\centering
    \includegraphics[width=0.99\linewidth]{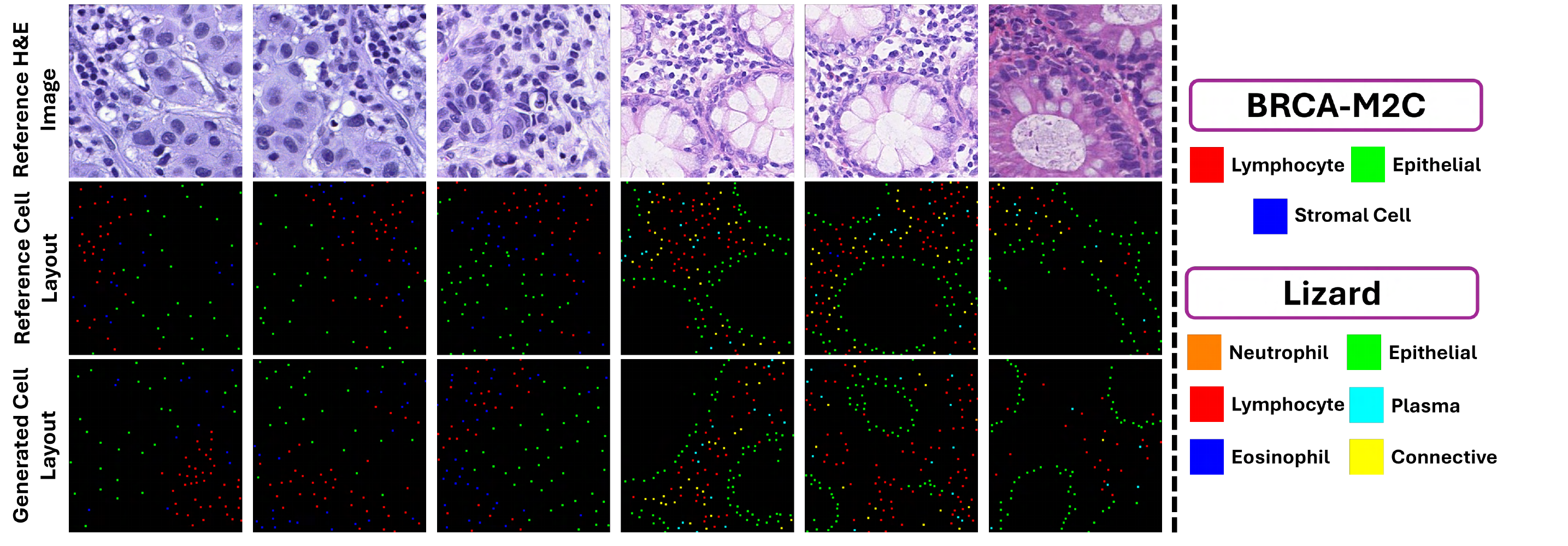}
    \vspace{-.1in}
    \caption{The qualitative results of our proposed \emph{TopoCellGen}. Columns $1$-$3$: BRCA-M2C dataset. Columns $4$-$6$: Lizard dataset. The cell types and their corresponding colors are shown on the right side of the figure.}
    \vspace{-.1in}
    \label{fig:qualitative_results}
\end{figure*}

To enable statistical analysis, we transform the persistence diagrams and their barycenters into persistence landscapes, transforming each diagram into a sequence of continuous, piecewise linear functions that preserve topological information while enabling standard statistical operations~\cite{bubenik2015statistical}. The vectorized barycenters are regarded as mean vectors $\boldsymbol{\mu}$ and we compute the covariance matrices $\boldsymbol{\Sigma}$ from both sets of persistence diagrams.

For multi-class layouts, TopoFD is calculated in two steps. First, for each cell type, we compute the vectorized persistence diagrams for both reference and synthetic layouts containing that type. Then, the final TopoFD is the average Fréchet Distance across cell types:
\begin{equation}
    \text{TopoFD} = \frac{1}{|n|} \sum_{i=1}^{n} \text{FD}(\boldsymbol{\mu}_i^r, \boldsymbol{\Sigma}_i^r, \boldsymbol{\mu}_i^s, \boldsymbol{\Sigma}_i^s)
\end{equation}
where $\text{FD}(\cdot)$ denotes the Fr\'echet distance computation.

\subsection{Layout-Guided Pathology Image Generation}
\label{subsec:layout-to-image}
In the final stage of our methodology, we focus on generating high-resolution pathology images based on the previously created cell layouts. These layouts serve as spatial and structural guides, ensuring that the pathology images adhere to the organization of cells within the layout. We employ a guided diffusion model~\cite{dhariwal2021diffusion} to synthesize realistic pathology images, conditioned on the underlying cell arrangement. This approach ensures that cells' spatial distribution and organization in the generated images align closely with the provided cell layouts. More details are provided in the Supplementary.

\section{Experiments}
We conduct extensive experiments on two public and widely used nuclei analysis datasets. We compare our method against SoTA layout generation methods in terms of sample quality and performance on downstream tasks.

\myparagraph{Datasets.}
We evaluate our proposed method on \textit{TCGA Breast Cancer Cell Classification Dataset (BRCA-M2C)}~\cite{abousamra2021multi} and \textit{Lizard dataset}~\cite{graham2021lizard}. More details about the datasets, the test set split, and the cell count distribution analysis are provided in the Supplementary.

\myparagraph{Evaluation Metrics.} We evaluate our proposed method on both sample quality and the performance of downstream tasks. We use \textbf{Fréchet Inception Distance (FID)}~\cite{heusel2017gans}, the \textbf{cell count error} for each cell type, the \textbf{total count error (TCE)}, our proposed \textbf{TopoFD} and \textbf{maximum mean discrepancy (MMD)}~\cite{wang2020topogan} to evaluate how well the generated cell layouts align with the reference layouts. 
Note that for FID, feature extraction is tailored to each dataset with custom-trained models.
On the other hand, we generate $2,000$ image-layout pairs as augmented training data for cell detection and classification tasks, evaluating their performance with the \textbf{F1-score}. More details are provided in the Supplementary.

\myparagraph{Implementation Details} are in the Supplementary.

\myparagraph{Biological Plausibility.} To further validate the quality of our generated cell layouts, we conducted an evaluation involving a board-certified pathologist with 7+ years of experience. From a pathologist’s perspective, the synthetic cell layouts are nearly indistinguishable from real ones and accurately capture the defining benign or malignant features of their real counterparts. More details are in the Supplementary.

\myparagraph{Backpropagation from Distance Transform Map to Binary Mask.}
To backpropagate the gradient from the distance transform map to a binary mask, we leverage an efficient implementation of the geodesic distance transform provided by the FastGeodis library~\cite{asad2022fastgeodis}. This method enables the computation of a differentiable distance map, which allows us to propagate gradients through the distance transform during training. The library implements a parallelized raster scan method that efficiently computes the Euclidean distance transforms on GPU hardware. 

\subsection{Experimental Results}

\myparagraph{Qualitative Results.}
Qualitative results are shown in \cref{fig:qualitative_results}. Our proposed \emph{TopoCellGen} method demonstrates cell layouts that closely match the reference layouts in both density and spatial arrangement. The generated layouts exhibit a high degree of distribution consistency, preserving density patterns and spatial relationships among cell types while reflecting the structural characteristics of the reference layouts.

\myparagraph{Quantitative Results on Sample Quality.}
~\cref{tab:sample_quality} shows the results of sample quality and generation accuracy on the BRCA-M2C and Lizard datasets compared with the state-of-the-art cell layout generation baselines: ADM ~\cite{dhariwal2021diffusion}, TMCCG ~\cite{abousamra2023topology} and Spatial Diffusion ~\cite{li2024spatial}. 
Across all multi-class datasets, our proposed \emph{TopoCellGen} attains the lowest FID and TopoFD, indicating superior visual fidelity and topological accuracy. Moreover, \emph{TopoCellGen} notably reduces multi-class count errors, resulting in more precise and topologically consistent synthetic cell layouts. These findings highlight \emph{TopoCellGen}’s capacity to preserve realistic cell distributions, maintain topological relationships, and enable accurate control over cell density.
\begin{table*}[ht]
\setlength{\tabcolsep}{5pt}
\centering
\vspace{-.1in}
\scriptsize
\begin{tabular}{ccccccccccccc}
\hline
 & Method & FID $\downarrow$ & Lym. $\downarrow$ & Epi. $\downarrow$ & Stro. $\downarrow$ & Neu. $\downarrow$ & Pla. $\downarrow$ & Eos. $\downarrow$ & Con. $\downarrow$ & TCE $\downarrow$ & TopoFD $\downarrow$ & MMD $\downarrow$ \\ \hline \hline
\multirow{4}{*}{BRCA-M2C} & ADM~\cite{dhariwal2021diffusion} & 1.150 & 13.757 & 40.230 & 15.491 & \multicolumn{1}{c}{--} & \multicolumn{1}{c}{--} & \multicolumn{1}{c}{--} & \multicolumn{1}{c}{--} & 22.465 & 133.012 & 0.732  \\ 
 & TMCCG~\cite{abousamra2023topology} & 0.634 & 11.503 & 34.032 & 12.907 & \multicolumn{1}{c}{--} & \multicolumn{1}{c}{--} & \multicolumn{1}{c}{--} & \multicolumn{1}{c}{--} & 19.687 & 89.252 & 0.635 \\ 
 & Spatial Diffusion~\cite{li2024spatial} & 0.263 &  10.852 & 35.954 & 13.496 & \multicolumn{1}{c}{--} & \multicolumn{1}{c}{--} & \multicolumn{1}{c}{--} & \multicolumn{1}{c}{--} & 20.806 & 97.584 & 0.589 \\ 
 & \emph{TopoCellGen} & \textbf{0.005} & \textbf{2.090} & \textbf{3.824} & \textbf{2.468} & \multicolumn{1}{c}{--} & \multicolumn{1}{c}{--} & \multicolumn{1}{c}{--} & \multicolumn{1}{c}{--} & \textbf{5.192} & \textbf{69.354} & \textbf{0.421} \\ \hline \hline 
 \multirow{4}{*}{Lizard} & ADM~\cite{dhariwal2021diffusion} & 0.059 & 16.508  & 11.796 & \multicolumn{1}{c}{--} & 1.123 & 4.328 & 1.598 & 10.737 & 23.964 & 65.910 & 0.783 \\ 
 & TMCCG~\cite{abousamra2023topology} & 1.093 & 15.548 & 10.011 & \multicolumn{1}{c}{--} &  2.376 & 4.293 & 1.872 & 11.643 & 22.604 & 63.120 &  0.667\\ 
 & Spatial Diffusion~\cite{li2024spatial}  & 0.137  & 10.740 & 9.062 & \multicolumn{1}{c}{--} & 3.040  & 6.552 & 2.173 & 11.225 & 20.606 & 79.591 & 0.883\\ 
 & \emph{TopoCellGen} & \textbf{0.027} & \textbf{6.155}  & \textbf{6.560} & \multicolumn{1}{c}{--} & \textbf{1.022} & \textbf{2.982} & \textbf{1.167} & \textbf{7.288} & \textbf{11.590} & \textbf{31.607} & \textbf{0.536} \\ \hline
\end{tabular}
\vspace{-.1in}
\caption{Results for BRCA-M2C and Lizard datasets on the quality of the generated samples.
}
\vspace{-.1in}
\label{tab:sample_quality}
\end{table*}

\begin{table*}[ht]
\centering
\scriptsize
\begin{tabular}{ccccccc}
\hline
\multirow{2}{*}{Data} & \multirow{2}{*}{Method} & \multicolumn{5}{c}{F1-Score $\uparrow$} \\ \cline{3-7} 
 &  & Lymphocytes & Epithelial & Stromal & Mean & Detection \\ \hline \hline
Real. & \multirow{5}{*}{UNet} & $0.569 \pm 0.010$ & $0.736 \pm 0.012$ & $0.507 \pm 0.015$ & $0.604 \pm 0.011$ & $0.857 \pm 0.006$ \\
Real+Syn. (Rand) & & $0.549 \pm 0.009$ & $0.693 \pm 0.014$ & $0.472 \pm 0.016$ & $0.571 \pm 0.013$ & $0.848 \pm 0.008$\\
Real+Syn (TMCCG) & & $0.650 \pm 0.007$ & $0.768 \pm 0.010$& $0.511 \pm 0.012$ & $0.643 \pm 0.009$& $0.852 \pm 0.005$ \\
Real+Syn (SpaDM) & & $0.647 \pm 0.006$ & $0.797 \pm 0.003$ & $0.554 \pm 0.011$& $0.666 \pm 0.007$ & $0.853 \pm 0.005$\\
Real+Syn (\emph{TopoCellGen}) &  & $\boldsymbol{0.656 \pm 0.003}$ & $\boldsymbol{0.803 \pm 0.005}$ & $\boldsymbol{0.574 \pm 0.004}$ & $\boldsymbol{0.678 \pm 0.004}$ & $\boldsymbol{0.860 \pm 0.004}$ \\ \hline \hline
Real. & \multirow{5}{*}{MCSpatNet} & $0.615 \pm 0.008$ & $0.777 \pm 0.010$ & $0.540 \pm 0.013$ & $0.644 \pm 0.009$ & $0.855 \pm 0.005$ \\
Real+Syn. (Rand) &  & $0.578 \pm 0.009$& $0.756 \pm 0.012$ & $0.502 \pm 0.014$ & $0.612 \pm 0.010$ & $0.851 \pm 0.006$ \\
Real+Syn (TMCCG) &  & $\boldsymbol{0.678 \pm 0.006}$ & $0.800 \pm 0.005$ & $0.522 \pm 0.014$ & $0.667 \pm 0.007$ & $0.853 \pm 0.004$ \\
Real+Syn (SpaDM) &  & $0.639 \pm 0.005$ & $0.804 \pm 0.007$ & $0.563 \pm 0.012$ & $0.669 \pm 0.006$ & $0.855 \pm 0.005$ \\
Real+Syn (\emph{TopoCellGen}) &  & $0.652 \pm 0.004$ & $\boldsymbol{0.817 \pm 0.006}$ & $\boldsymbol{0.582 \pm 0.005}$ & $\boldsymbol{0.684 \pm 0.004}$ & $\boldsymbol{0.862 \pm 0.004}$ \\ \hline
\end{tabular}
\vspace{-.1in}
\caption{Results on cell detection and classification tasks on BRCA-M2C dataset. The best and statistically significant results are highlighted in \textbf{bold}.
}
\vspace{-.2in}
\label{tab:downstream_tasks}
\end{table*}

\myparagraph{Performance on Downstream Tasks. }In~\cref{tab:downstream_tasks}, we show the results of cell detection and classification tasks using our synthetic image-layout pairs as data augmentations. We present the results using two frameworks, UNet~\cite{ronneberger2015u} and MCSpatNet~\cite{abousamra2021multi}. The results indicate that \emph{TopoCellGen} achieves the highest F1 scores across various cell types, including inflammation and epithelial cells, resulting in superior mean F1 scores and detection metrics on the BRCA-M2C dataset. This demonstrates our \emph{TopoCellGen}'s capability to capture both spatial fidelity and topological accuracy, ensuring that synthetic data closely resembles real biological structures. By accurately modeling complex spatial distributions and inter-class relationships, \emph{TopoCellGen} provides biologically plausible synthetic samples that improve the generalizability of detection and classification models. Furthermore, its balanced representation of cellular compositions reduces class-wise biases, allowing the classifiers to better learn fine-grained distinctions.

\subsection{Ablation Studies}
Extensive experiments are conducted to elucidate the effectiveness and robustness of our loss components and hyper-parameters. All experiments are performed on the BRCA-M2C dataset.

\myparagraph{Ablation Study on Loss Components.} 
We evaluate the contributions of three loss functions to the model's performance in generating synthetic cell layouts, measured by FID, Total Counting Error, and TopoFD. As shown in \cref{ablation:loss_components}, the results demonstrate that $\mathcal{L}_{\text{count}}$ significantly reduces the counting errors, achieving better accuracy when combined with the other losses. The $\mathcal{L}_{\text{intra}}$ improves the fidelity and topological accuracy of the generated multi-class layouts, as indicated by lower FID and TopoFD. The $\mathcal{L}_{\text{inter}}$ has a smaller effect on the cell counting error but enhances structural consistency, as reflected in improved TopoFD. 
Combining all three losses yields superior results across all metrics, highlighting their complementary contributions to accurate and realistic multi-class cell layouts.
\begin{table}[ht]
\centering
\scriptsize
\begin{tabular}{cccccc}
\hline
\multirow{2}{*}{$\mathcal{L}_{\text{count}}$} & \multirow{2}{*}{$\mathcal{L}_{\text{intra}}$} & \multirow{2}{*}{$\mathcal{L}_{\text{inter}}$} & \multicolumn{3}{c}{BRCA-M2C} \\ \cline{4-6} 
 &  &  & FID $\downarrow$ & TCE $\downarrow$& TopoFD $\downarrow$\\ \hline \hline
 \xmark & \xmark & \xmark & 1.150 & 22.465 & 133.012 \\
 \cmark & \xmark & \xmark & 0.842 & 12.253 & 118.304 \\
 \xmark & \cmark & \xmark & 0.621 & 16.315 & 98.798 \\
 \xmark & \xmark & \cmark & 1.083 & 21.928 & 126.742 \\
 \cmark & \cmark & \xmark & 0.232 & 6.854 & 85.672 \\
 \cmark & \xmark & \cmark & 0.498 & 8.012 & 91.324 \\
 \xmark & \cmark & \cmark & 0.327 & 17.573 & 73.612 \\
 \cmark & \cmark & \cmark & \textbf{0.005} & \textbf{5.192} & \textbf{69.354} \\ \hline
\end{tabular}
\vspace{-.1in}
\caption{Ablation study on loss components.}
\vspace{-.2in}
\label{ablation:loss_components}
\end{table}

\myparagraph{Ablation Study on Loss Weights.}
The ablation study on loss weights is presented in \cref{ablation:loss_weights}. The results indicate that when the loss weights are higher, such as $0.005$ or $0.001$, they impose overly strong regularization, leading to suboptimal outcomes. Conversely, when the loss weights are lowered to $1e-4$ or even $5e-5$, the regularization becomes too weak to achieve optimal performance. The configuration with moderate weights $5e-4$ achieves the best balance, yielding the lowest FID, total count error, and TopoFD. 
\begin{table}[ht]
\centering
\scriptsize
\begin{tabular}{cccccc}
\hline
\multirow{2}{*}{$\lambda_{c}$} & \multirow{2}{*}{$\lambda_{\text{intra}}$} & \multirow{2}{*}{$\lambda_{\text{inter}}$} & \multicolumn{3}{c}{BRCA-M2C} \\ \cline{4-6} 
 &  &  & FID $\downarrow$ & TCE $\downarrow$ & TopoFD $\downarrow$\\ \hline \hline
 0.005 & 0.005 & 0.005 & 0.289 & 14.920 & 92.718 \\
 0.001 & 0.001 & 0.001 & 0.153 & 12.471 & 85.642 \\
 1e-4 & 1e-4 & 1e-4 & 0.012 & 8.275 & 88.627 \\
 5e-5 & 5e-5 & 5e-5 & 0.129 & 11.378 & 83.629 \\
 5e-4 & 5e-4 & 5e-4 & \textbf{0.005} & \textbf{5.192} & \textbf{69.354} \\ \hline
\end{tabular}
\caption{Ablation study on loss weights.}
\label{ablation:loss_weights}
\vspace{-.2in}
\end{table}

\section{Conclusion}
In summary, \emph{TopoCellGen} presents a robust framework for generating realistic cell topologies in digital pathology. It accurately preserves both intra- and inter-class spatial patterns, ensures cell count control, and achieves high structural fidelity. Experimental results confirm its close approximation of real tissue layouts, thereby enhancing downstream tasks such as cell detection and classification.

\myparagraph{Acknowledgment.} We thank Dr.~Michael L.~Miller for evaluating the biological plausibility of our results.
This research was partially supported by the National Science Foundation (NSF) grants CCF-2144901, IIS-2123920, IIS-2212046, the National Institute of Health (NIH) grants R01GM148970, R01GM148970-03S1, R21CA258493, R21CA258493-01A1, R01CA297843, and the Stony Brook Trustees Faculty Award.

{
    \small
    \bibliographystyle{ieeenat_fullname}
    \bibliography{main}

\begin{thebibliography}{77}
\providecommand{\natexlab}[1]{#1}
\providecommand{\url}[1]{\texttt{#1}}
\expandafter\ifx\csname urlstyle\endcsname\relax
  \providecommand{\doi}[1]{doi: #1}\else
  \providecommand{\doi}{doi: \begingroup \urlstyle{rm}\Url}\fi

\bibitem[Abousamra et~al.(2021)Abousamra, Belinsky, Van~Arnam, Allard, Yee, Gupta, Kurc, Samaras, Saltz, and Chen]{abousamra2021multi}
Shahira Abousamra, David Belinsky, John Van~Arnam, Felicia Allard, Eric Yee, Rajarsi Gupta, Tahsin Kurc, Dimitris Samaras, Joel Saltz, and Chao Chen.
\newblock Multi-class cell detection using spatial context representation.
\newblock In \emph{ICCV}, 2021.

\bibitem[Abousamra et~al.(2023)Abousamra, Gupta, Kurc, Samaras, Saltz, and Chen]{abousamra2023topology}
Shahira Abousamra, Rajarsi Gupta, Tahsin Kurc, Dimitris Samaras, Joel Saltz, and Chao Chen.
\newblock Topology-guided multi-class cell context generation for digital pathology.
\newblock In \emph{CVPR}, 2023.

\bibitem[Asad et~al.(2022)Asad, Dorent, and Vercauteren]{asad2022fastgeodis}
Muhammad Asad, Reuben Dorent, and Tom Vercauteren.
\newblock Fastgeodis: Fast generalised geodesic distance transform.
\newblock \emph{Journal of Open Source Software}, 2022.

\bibitem[Aversa et~al.(2024)Aversa, Nobis, H{\"a}gele, Standvoss, Chirica, Murray-Smith, Alaa, Ruff, Ivanova, Samek, et~al.]{aversa2024diffinfinite}
Marco Aversa, Gabriel Nobis, Miriam H{\"a}gele, Kai Standvoss, Mihaela Chirica, Roderick Murray-Smith, Ahmed~M Alaa, Lukas Ruff, Daniela Ivanova, Wojciech Samek, et~al.
\newblock Diffinfinite: Large mask-image synthesis via parallel random patch diffusion in histopathology.
\newblock \emph{NeurIPS}, 2024.

\bibitem[Bengio et~al.(2013)Bengio, L{\'e}onard, and Courville]{bengio2013estimating}
Yoshua Bengio, Nicholas L{\'e}onard, and Aaron Courville.
\newblock Estimating or propagating gradients through stochastic neurons for conditional computation.
\newblock \emph{arXiv preprint arXiv:1308.3432}, 2013.

\bibitem[Brooks et~al.(2023)Brooks, Holynski, and Efros]{brooks2023instructpix2pix}
Tim Brooks, Aleksander Holynski, and Alexei~A Efros.
\newblock Instructpix2pix: Learning to follow image editing instructions.
\newblock In \emph{CVPR}, 2023.

\bibitem[Bubenik et~al.(2015)]{bubenik2015statistical}
Peter Bubenik et~al.
\newblock Statistical topological data analysis using persistence landscapes.
\newblock \emph{JMLR}, 2015.

\bibitem[Clough et~al.(2020)Clough, Byrne, Oksuz, Zimmer, Schnabel, and King]{clough2020topological}
James~R Clough, Nicholas Byrne, Ilkay Oksuz, Veronika~A Zimmer, Julia~A Schnabel, and Andrew~P King.
\newblock A topological loss function for deep-learning based image segmentation using persistent homology.
\newblock \emph{TPAMI}, 2020.

\bibitem[Cohen-Steiner et~al.(2010)Cohen-Steiner, Edelsbrunner, Harer, and Mileyko]{cohen2010lipschitz}
David Cohen-Steiner, Herbert Edelsbrunner, John Harer, and Yuriy Mileyko.
\newblock Lipschitz functions have l p-stable persistence.
\newblock \emph{Foundations of computational mathematics}, 2010.

\bibitem[Deshpande et~al.(2024)Deshpande, Dawood, Minhas, and Rajpoot]{deshpande2024synclay}
Srijay Deshpande, Muhammad Dawood, Fayyaz Minhas, and Nasir Rajpoot.
\newblock Synclay: Interactive synthesis of histology images from bespoke cellular layouts.
\newblock \emph{Medical Image Analysis}, 2024.

\bibitem[Dhariwal and Nichol(2021)]{dhariwal2021diffusion}
Prafulla Dhariwal and Alexander Nichol.
\newblock Diffusion models beat gans on image synthesis.
\newblock In \emph{NeurIPS}, 2021.

\bibitem[Ding et~al.(2022)Ding, Prasanna, Corredor, Barrera, Zens, Lu, Velu, Leo, Beig, Li, et~al.]{ding2022image}
Ruiwen Ding, Prateek Prasanna, Germ{\'a}n Corredor, Cristian Barrera, Philipp Zens, Cheng Lu, Priya Velu, Patrick Leo, Niha Beig, Haojia Li, et~al.
\newblock Image analysis reveals molecularly distinct patterns of tils in nsclc associated with treatment outcome.
\newblock \emph{NPJ precision oncology}, 2022.

\bibitem[Edelsbrunner et~al.(2002)Edelsbrunner, Letscher, and Zomorodian]{edelsbrunner2002topological}
Edelsbrunner, Letscher, and Zomorodian.
\newblock Topological persistence and simplification.
\newblock \emph{Discrete \& Computational Geometry}, 2002.

\bibitem[Edelsbrunner and Harer(2022)]{edelsbrunner2022computational}
Herbert Edelsbrunner and John~L Harer.
\newblock \emph{Computational topology: an introduction}.
\newblock American Mathematical Society, 2022.

\bibitem[Fei et~al.(2023)Fei, Lyu, Pan, Zhang, Yang, Luo, Zhang, and Dai]{fei2023generative}
Ben Fei, Zhaoyang Lyu, Liang Pan, Junzhe Zhang, Weidong Yang, Tianyue Luo, Bo Zhang, and Bo Dai.
\newblock Generative diffusion prior for unified image restoration and enhancement.
\newblock In \emph{CVPR}, 2023.

\bibitem[Felzenszwalb and Huttenlocher(2012)]{felzenszwalb2012distance}
Pedro~F Felzenszwalb and Daniel~P Huttenlocher.
\newblock Distance transforms of sampled functions.
\newblock \emph{Theory of computing}, 2012.

\bibitem[Forman(1998)]{forman1998morse}
Robin Forman.
\newblock Morse theory for cell complexes.
\newblock \emph{Advances in mathematics}, 1998.

\bibitem[Fr{\'e}chet(1957)]{frechet1957distance}
Maurice Fr{\'e}chet.
\newblock Sur la distance de deux lois de probabilit{\'e}.
\newblock In \emph{Annales de l'ISUP}, 1957.

\bibitem[Gabrielsson et~al.(2020)Gabrielsson, Nelson, Dwaraknath, and Skraba]{gabrielsson2020topology}
Rickard~Br{\"u}el Gabrielsson, Bradley~J Nelson, Anjan Dwaraknath, and Primoz Skraba.
\newblock A topology layer for machine learning.
\newblock In \emph{AISTATS}, 2020.

\bibitem[Goodfellow et~al.(2020)Goodfellow, Pouget-Abadie, Mirza, Xu, Warde-Farley, Ozair, Courville, and Bengio]{goodfellow2020generative}
Ian Goodfellow, Jean Pouget-Abadie, Mehdi Mirza, Bing Xu, David Warde-Farley, Sherjil Ozair, Aaron Courville, and Yoshua Bengio.
\newblock Generative adversarial networks.
\newblock \emph{Communications of the ACM}, 2020.

\bibitem[Graham et~al.(2019)Graham, Vu, Raza, Azam, Tsang, Kwak, and Rajpoot]{graham2019hover}
Simon Graham, Quoc~Dang Vu, Shan E~Ahmed Raza, Ayesha Azam, Yee~Wah Tsang, Jin~Tae Kwak, and Nasir Rajpoot.
\newblock Hover-net: Simultaneous segmentation and classification of nuclei in multi-tissue histology images.
\newblock \emph{MedIA}, 2019.

\bibitem[Graham et~al.(2021)Graham, Jahanifar, Azam, Nimir, Tsang, Dodd, Hero, Sahota, Tank, Benes, et~al.]{graham2021lizard}
Simon Graham, Mostafa Jahanifar, Ayesha Azam, Mohammed Nimir, Yee-Wah Tsang, Katherine Dodd, Emily Hero, Harvir Sahota, Atisha Tank, Ksenija Benes, et~al.
\newblock Lizard: A large-scale dataset for colonic nuclear instance segmentation and classification.
\newblock In \emph{ICCV}, 2021.

\bibitem[Graikos et~al.(2024)Graikos, Yellapragada, Le, Kapse, Prasanna, Saltz, and Samaras]{graikos2024learned}
Alexandros Graikos, Srikar Yellapragada, Minh-Quan Le, Saarthak Kapse, Prateek Prasanna, Joel Saltz, and Dimitris Samaras.
\newblock Learned representation-guided diffusion models for large-image generation.
\newblock In \emph{CVPR}, 2024.

\bibitem[Gretton et~al.(2012)Gretton, Borgwardt, Rasch, Sch{\"o}lkopf, and Smola]{gretton2012kernel}
Arthur Gretton, Karsten~M Borgwardt, Malte~J Rasch, Bernhard Sch{\"o}lkopf, and Alexander Smola.
\newblock A kernel two-sample test.
\newblock \emph{JMLR}, 2012.

\bibitem[Gupta et~al.(2022)Gupta, Hu, Kaan, Jin, Mpoy, Chung, Singh, Saltz, Kurc, Saltz, et~al.]{gupta2022learning}
Saumya Gupta, Xiaoling Hu, James Kaan, Michael Jin, Mutshipay Mpoy, Katherine Chung, Gagandeep Singh, Mary Saltz, Tahsin Kurc, Joel Saltz, et~al.
\newblock Learning topological interactions for multi-class medical image segmentation.
\newblock In \emph{ECCV}, 2022.

\bibitem[Gupta et~al.(2024{\natexlab{a}})Gupta, Samaras, and Chen]{gupta2024topodiffusionnet}
Saumya Gupta, Dimitris Samaras, and Chao Chen.
\newblock Topodiffusionnet: A topology-aware diffusion model.
\newblock \emph{arXiv preprint arXiv:2410.16646}, 2024{\natexlab{a}}.

\bibitem[Gupta et~al.(2024{\natexlab{b}})Gupta, Zhang, Hu, Prasanna, and Chen]{gupta2024topology}
Saumya Gupta, Yikai Zhang, Xiaoling Hu, Prateek Prasanna, and Chao Chen.
\newblock Topology-aware uncertainty for image segmentation.
\newblock In \emph{NeurIPS}, 2024{\natexlab{b}}.

\bibitem[Harb et~al.(2024)Harb, Pock, and M{\"u}ller]{harb2024diffusion}
Robert Harb, Thomas Pock, and Heimo M{\"u}ller.
\newblock Diffusion-based generation of histopathological whole slide images at a gigapixel scale.
\newblock In \emph{WACV}, 2024.

\bibitem[He et~al.(2023)He, Wang, Wei, Xu, Ji, Liu, and Chen]{he2023toposeg}
Hongliang He, Jun Wang, Pengxu Wei, Fan Xu, Xiangyang Ji, Chang Liu, and Jie Chen.
\newblock Toposeg: Topology-aware nuclear instance segmentation.
\newblock In \emph{ICCV}, 2023.

\bibitem[Heusel et~al.(2017)Heusel, Ramsauer, Unterthiner, Nessler, and Hochreiter]{heusel2017gans}
Martin Heusel, Hubert Ramsauer, Thomas Unterthiner, Bernhard Nessler, and Sepp Hochreiter.
\newblock Gans trained by a two time-scale update rule converge to a local nash equilibrium.
\newblock In \emph{NeurIPS}, 2017.

\bibitem[Ho et~al.(2020)Ho, Jain, and Abbeel]{ho2020denoising}
Jonathan Ho, Ajay Jain, and Pieter Abbeel.
\newblock Denoising diffusion probabilistic models.
\newblock In \emph{NeurIPS}, 2020.

\bibitem[H{\"o}rst et~al.(2024)H{\"o}rst, Rempe, Heine, Seibold, Keyl, Baldini, Ugurel, Siveke, Gr{\"u}nwald, Egger, et~al.]{horst2024cellvit}
Fabian H{\"o}rst, Moritz Rempe, Lukas Heine, Constantin Seibold, Julius Keyl, Giulia Baldini, Selma Ugurel, Jens Siveke, Barbara Gr{\"u}nwald, Jan Egger, et~al.
\newblock Cellvit: Vision transformers for precise cell segmentation and classification.
\newblock \emph{MedIA}, 2024.

\bibitem[Hou et~al.(2024)Hou, Zhu, Hou, Liu, Zeng, and Yuan]{hou2024global}
Jinhui Hou, Zhiyu Zhu, Junhui Hou, Hui Liu, Huanqiang Zeng, and Hui Yuan.
\newblock Global structure-aware diffusion process for low-light image enhancement.
\newblock In \emph{NeurIPS}, 2024.

\bibitem[Hou et~al.(2019)Hou, Agarwal, Samaras, Kurc, Gupta, and Saltz]{hou2019robust}
Le Hou, Ayush Agarwal, Dimitris Samaras, Tahsin~M Kurc, Rajarsi~R Gupta, and Joel~H Saltz.
\newblock Robust histopathology image analysis: To label or to synthesize?
\newblock In \emph{CVPR}, 2019.

\bibitem[Hu(2022)]{hu2022structure}
Xiaoling Hu.
\newblock Structure-aware image segmentation with homotopy warping.
\newblock In \emph{NeurIPS}, 2022.

\bibitem[Hu et~al.(2019)Hu, Li, Samaras, and Chen]{hu2019topology}
Xiaoling Hu, Fuxin Li, Dimitris Samaras, and Chao Chen.
\newblock Topology-preserving deep image segmentation.
\newblock In \emph{NeurIPS}, 2019.

\bibitem[Hu et~al.(2021)Hu, Wang, Fuxin, Samaras, and Chen]{hu2020topology}
Xiaoling Hu, Yusu Wang, Li Fuxin, Dimitris Samaras, and Chao Chen.
\newblock Topology-aware segmentation using discrete morse theory.
\newblock In \emph{ICLR}, 2021.

\bibitem[Hu et~al.(2023)Hu, Samaras, and Chen]{hu2022learning}
Xiaoling Hu, Dimitris Samaras, and Chao Chen.
\newblock Learning probabilistic topological representations using discrete morse theory.
\newblock In \emph{ICLR}, 2023.

\bibitem[Kerber et~al.(2016)Kerber, Morozov, and Nigmetov]{kerber2016geometry}
Michael Kerber, Dmitriy Morozov, and Arnur Nigmetov.
\newblock Geometry helps to compare persistence diagrams.
\newblock In \emph{2016 Proceedings of the Eighteenth Workshop on Algorithm Engineering and Experiments (ALENEX)}, 2016.

\bibitem[Koohbanani et~al.(2020)Koohbanani, Jahanifar, Tajadin, and Rajpoot]{koohbanani2020nuclick}
Navid~Alemi Koohbanani, Mostafa Jahanifar, Neda~Zamani Tajadin, and Nasir Rajpoot.
\newblock Nuclick: a deep learning framework for interactive segmentation of microscopic images.
\newblock \emph{MedIA}, 2020.

\bibitem[Lacombe et~al.(2018)Lacombe, Cuturi, and Oudot]{lacombe2018large}
Th{\'e}o Lacombe, Marco Cuturi, and Steve Oudot.
\newblock Large scale computation of means and clusters for persistence diagrams using optimal transport.
\newblock In \emph{NeurIPS}, 2018.

\bibitem[Lee et~al.(2008)Lee, Park, Kay, Christakis, Oltvai, and Barab{\'a}si]{lee2008implications}
D-S Lee, Juyong Park, KA Kay, Nicholas~A Christakis, Zoltan~N Oltvai, and A-L Barab{\'a}si.
\newblock The implications of human metabolic network topology for disease comorbidity.
\newblock \emph{PNAS}, 2008.

\bibitem[Levenson et~al.(2024)Levenson, Singh, Rieck, Hathaway, Farrelly, Rozenblit, Prasanna, Erickson, Choudhary, Carlsson, et~al.]{levenson2024advancing}
Richard~M Levenson, Yashbir Singh, Bastian Rieck, Quincy~A Hathaway, Colleen Farrelly, Jennifer Rozenblit, Prateek Prasanna, Bradley Erickson, Ashok Choudhary, Gunnar Carlsson, et~al.
\newblock Advancing precision medicine: algebraic topology and differential geometry in radiology and computational pathology.
\newblock \emph{Laboratory Investigation}, 2024.

\bibitem[Li et~al.(2024{\natexlab{a}})Li, Hu, Abousamra, Xu, and Chen]{li2024spatial}
Chen Li, Xiaoling Hu, Shahira Abousamra, Meilong Xu, and Chao Chen.
\newblock Spatial diffusion for cell layout generation.
\newblock In \emph{MICCAI}, 2024{\natexlab{a}}.

\bibitem[Li et~al.(2024{\natexlab{b}})Li, Yang, Kuang, Wu, Wang, Xiao, and Chen]{li2025controlnet}
Ming Li, Taojiannan Yang, Huafeng Kuang, Jie Wu, Zhaoning Wang, Xuefeng Xiao, and Chen Chen.
\newblock Controlnet++: Improving conditional controls with efficient consistency feedback.
\newblock In \emph{ECCV}, 2024{\natexlab{b}}.

\bibitem[Lu et~al.(2021)Lu, Koyuncu, Corredor, Prasanna, Leo, Wang, Janowczyk, Bera, Lewis~Jr, Velcheti, et~al.]{lu2021feature}
Cheng Lu, Can Koyuncu, German Corredor, Prateek Prasanna, Patrick Leo, XiangXue Wang, Andrew Janowczyk, Kaustav Bera, James Lewis~Jr, Vamsidhar Velcheti, et~al.
\newblock Feature-driven local cell graph (flock): new computational pathology-based descriptors for prognosis of lung cancer and hpv status of oropharyngeal cancers.
\newblock \emph{MedIA}, 2021.

\bibitem[Lugli et~al.(2021)Lugli, Zlobec, Berger, Kirsch, and Nagtegaal]{lugli2021tumour}
Alessandro Lugli, Inti Zlobec, Martin~D Berger, Richard Kirsch, and Iris~D Nagtegaal.
\newblock Tumour budding in solid cancers.
\newblock \emph{Nature Reviews Clinical Oncology}, 2021.

\bibitem[Min et~al.(2024)Min, Oh, and Jeong]{min2024co}
Seonghui Min, Hyun-Jic Oh, and Won-Ki Jeong.
\newblock Co-synthesis of histopathology nuclei image-label pairs using a context-conditioned joint diffusion model.
\newblock In \emph{ECCV}, 2024.

\bibitem[Munkres(1984)]{munkres1984elements}
James~R Munkres.
\newblock Elements of algebraic topology, 1984.

\bibitem[Nichol and Dhariwal(2021)]{nichol2021improved}
Alexander~Quinn Nichol and Prafulla Dhariwal.
\newblock Improved denoising diffusion probabilistic models.
\newblock In \emph{ICML}, 2021.

\bibitem[Oh and Jeong(2023)]{oh2023diffmix}
Hyun-Jic Oh and Won-Ki Jeong.
\newblock Diffmix: Diffusion model-based data synthesis for nuclei segmentation and classification in imbalanced pathology image datasets.
\newblock In \emph{MICCAI}, 2023.

\bibitem[Prangemeier et~al.(2020)Prangemeier, Reich, and Koeppl]{prangemeier2020attention}
Tim Prangemeier, Christoph Reich, and Heinz Koeppl.
\newblock Attention-based transformers for instance segmentation of cells in microstructures.
\newblock In \emph{2020 IEEE international conference on Bioinformatics and Biomedicine (BIBM)}. IEEE, 2020.

\bibitem[Rombach et~al.(2022)Rombach, Blattmann, Lorenz, Esser, and Ommer]{rombach2022high}
Robin Rombach, Andreas Blattmann, Dominik Lorenz, Patrick Esser, and Bj{\"o}rn Ommer.
\newblock High-resolution image synthesis with latent diffusion models.
\newblock In \emph{CVPR}, 2022.

\bibitem[Ronneberger et~al.(2015)Ronneberger, Fischer, and Brox]{ronneberger2015u}
Olaf Ronneberger, Philipp Fischer, and Thomas Brox.
\newblock U-net: Convolutional networks for biomedical image segmentation.
\newblock In \emph{MICCAI}, 2015.

\bibitem[Saharia et~al.(2022)Saharia, Chan, Saxena, Li, Whang, Denton, Ghasemipour, Gontijo~Lopes, Karagol~Ayan, Salimans, et~al.]{saharia2022photorealistic}
Chitwan Saharia, William Chan, Saurabh Saxena, Lala Li, Jay Whang, Emily~L Denton, Kamyar Ghasemipour, Raphael Gontijo~Lopes, Burcu Karagol~Ayan, Tim Salimans, et~al.
\newblock Photorealistic text-to-image diffusion models with deep language understanding.
\newblock In \emph{NeurIPS}, 2022.

\bibitem[Salgado et~al.(2015)Salgado, Denkert, Demaria, Sirtaine, Klauschen, Pruneri, Wienert, Van~den Eynden, Baehner, P{\'e}nault-Llorca, et~al.]{salgado2015evaluation}
Roberto Salgado, Carsten Denkert, S Demaria, N Sirtaine, F Klauschen, Giancarlo Pruneri, S Wienert, Gert Van~den Eynden, Frederick~L Baehner, Frederique P{\'e}nault-Llorca, et~al.
\newblock The evaluation of tumor-infiltrating lymphocytes (tils) in breast cancer: recommendations by an international tils working group 2014.
\newblock \emph{Annals of oncology}, 2015.

\bibitem[Saltz et~al.(2018)Saltz, Gupta, Hou, Kurc, Singh, Nguyen, Samaras, Shroyer, Zhao, Batiste, et~al.]{saltz2018spatial}
Joel Saltz, Rajarsi Gupta, Le Hou, Tahsin Kurc, Pankaj Singh, Vu Nguyen, Dimitris Samaras, Kenneth~R Shroyer, Tianhao Zhao, Rebecca Batiste, et~al.
\newblock Spatial organization and molecular correlation of tumor-infiltrating lymphocytes using deep learning on pathology images.
\newblock \emph{Cell reports}, 2018.

\bibitem[Samet and Tamminen(1988)]{samet1988efficient}
Hanan Samet and Markku Tamminen.
\newblock Efficient component labeling of images of arbitrary dimension represented by linear bintrees.
\newblock \emph{TPAMI}, 1988.

\bibitem[Schapiro et~al.(2017)Schapiro, Jackson, Raghuraman, Fischer, Zanotelli, Schulz, Giesen, Catena, Varga, and Bodenmiller]{schapiro2017histocat}
Denis Schapiro, Hartland~W Jackson, Swetha Raghuraman, Jana~R Fischer, Vito~RT Zanotelli, Daniel Schulz, Charlotte Giesen, Ra{\'u}l Catena, Zsuzsanna Varga, and Bernd Bodenmiller.
\newblock histocat: analysis of cell phenotypes and interactions in multiplex image cytometry data.
\newblock \emph{Nature methods}, 2017.

\bibitem[Shit et~al.(2021)Shit, Paetzold, Sekuboyina, Ezhov, Unger, Zhylka, Pluim, Bauer, and Menze]{shit2021cldice}
Suprosanna Shit, Johannes~C Paetzold, Anjany Sekuboyina, Ivan Ezhov, Alexander Unger, Andrey Zhylka, Josien~PW Pluim, Ulrich Bauer, and Bjoern~H Menze.
\newblock cldice-a novel topology-preserving loss function for tubular structure segmentation.
\newblock In \emph{CVPR}, 2021.

\bibitem[Sohl-Dickstein et~al.(2015)Sohl-Dickstein, Weiss, Maheswaranathan, and Ganguli]{sohl2015deep}
Jascha Sohl-Dickstein, Eric Weiss, Niru Maheswaranathan, and Surya Ganguli.
\newblock Deep unsupervised learning using nonequilibrium thermodynamics.
\newblock In \emph{ICML}, 2015.

\bibitem[Song et~al.(2021)Song, Meng, and Ermon]{song2020denoising}
Jiaming Song, Chenlin Meng, and Stefano Ermon.
\newblock Denoising diffusion implicit models.
\newblock In \emph{ICLR}, 2021.

\bibitem[Stanton and Disis(2016)]{stanton2016clinical}
Sasha~E Stanton and Mary~L Disis.
\newblock Clinical significance of tumor-infiltrating lymphocytes in breast cancer.
\newblock \emph{Journal for immunotherapy of cancer}, 2016.

\bibitem[Stucki et~al.(2023)Stucki, Paetzold, Shit, Menze, and Bauer]{stucki2023topologically}
Nico Stucki, Johannes~C Paetzold, Suprosanna Shit, Bjoern Menze, and Ulrich Bauer.
\newblock Topologically faithful image segmentation via induced matching of persistence barcodes.
\newblock In \emph{ICML}, 2023.

\bibitem[Szegedy et~al.(2016)Szegedy, Vanhoucke, Ioffe, Shlens, and Wojna]{szegedy2016rethinking}
Christian Szegedy, Vincent Vanhoucke, Sergey Ioffe, Jon Shlens, and Zbigniew Wojna.
\newblock Rethinking the inception architecture for computer vision.
\newblock In \emph{CVPR}, 2016.

\bibitem[Wang et~al.(2020)Wang, Liu, Samaras, and Chen]{wang2020topogan}
Fan Wang, Huidong Liu, Dimitris Samaras, and Chao Chen.
\newblock Topogan: A topology-aware generative adversarial network.
\newblock In \emph{ECCV}, 2020.

\bibitem[Wang et~al.(2024)Wang, Zou, Sakla, Partyka, Rawal, Singh, Zhao, Ling, Huang, Prasanna, et~al.]{wang2024topotxr}
Fan Wang, Zhilin Zou, Nicole Sakla, Luke Partyka, Nil Rawal, Gagandeep Singh, Wei Zhao, Haibin Ling, Chuan Huang, Prateek Prasanna, et~al.
\newblock Topotxr: A topology-guided deep convolutional network for breast parenchyma learning on dce-mris.
\newblock \emph{MedIA}, 2024.

\bibitem[Wang et~al.(2022)Wang, Xian, and Vakanski]{wang2022ta}
Haotian Wang, Min Xian, and Aleksandar Vakanski.
\newblock Ta-net: Topology-aware network for gland segmentation.
\newblock In \emph{WACV}, 2022.

\bibitem[Wang et~al.(2023)Wang, Levman, Pinaya, Tudosiu, Cardoso, and Marinescu]{wang2023inversesr}
Jueqi Wang, Jacob Levman, Walter Hugo~Lopez Pinaya, Petru-Daniel Tudosiu, M~Jorge Cardoso, and Razvan Marinescu.
\newblock Inversesr: 3d brain mri super-resolution using a latent diffusion model.
\newblock In \emph{MICCAI}. Springer, 2023.

\bibitem[Xu et~al.(2024{\natexlab{a}})Xu, Gao, Chen, Garai, Duong-Tran, Zhao, and Shen]{xu2024topology}
Frederick~H Xu, Michael Gao, Jiong Chen, Sumita Garai, Duy~Anh Duong-Tran, Yize Zhao, and Li Shen.
\newblock Topology-based clustering of functional brain networks in an alzheimer’s disease cohort.
\newblock \emph{AMIA Summits on Translational Science Proceedings}, 2024{\natexlab{a}}.

\bibitem[Xu et~al.(2024{\natexlab{b}})Xu, Hu, Gupta, Abousamra, and Chen]{xu2025semi}
Meilong Xu, Xiaoling Hu, Saumya Gupta, Shahira Abousamra, and Chao Chen.
\newblock Semi-supervised segmentation of histopathology images with noise-aware topological consistency.
\newblock In \emph{ECCV}, 2024{\natexlab{b}}.

\bibitem[Yellapragada et~al.(2024)Yellapragada, Graikos, Prasanna, Kurc, Saltz, and Samaras]{yellapragada2024pathldm}
Srikar Yellapragada, Alexandros Graikos, Prateek Prasanna, Tahsin Kurc, Joel Saltz, and Dimitris Samaras.
\newblock Pathldm: Text conditioned latent diffusion model for histopathology.
\newblock In \emph{WACV}, 2024.

\bibitem[You et~al.(2024)You, Dai, Liu, Min, Dvornek, Li, Clifton, Staib, and Duncan]{you2024mine}
Chenyu You, Weicheng Dai, Fenglin Liu, Yifei Min, Nicha~C Dvornek, Xiaoxiao Li, David~A Clifton, Lawrence Staib, and James~S Duncan.
\newblock Mine your own anatomy: Revisiting medical image segmentation with extremely limited labels.
\newblock \emph{TPAMI}, 2024.

\bibitem[Yuan(2016)]{yuan2016spatial}
Yinyin Yuan.
\newblock Spatial heterogeneity in the tumor microenvironment.
\newblock \emph{Cold Spring Harbor perspectives in medicine}, 2016.

\bibitem[Zhang et~al.(2023)Zhang, Rao, and Agrawala]{zhang2023adding}
Lvmin Zhang, Anyi Rao, and Maneesh Agrawala.
\newblock Adding conditional control to text-to-image diffusion models.
\newblock In \emph{ICCV}, 2023.

\bibitem[Zhao et~al.(2024)Zhao, Chen, Chen, Bao, Hao, Yuan, and Wong]{zhao2024uni}
Shihao Zhao, Dongdong Chen, Yen-Chun Chen, Jianmin Bao, Shaozhe Hao, Lu Yuan, and Kwan-Yee~K Wong.
\newblock Uni-controlnet: All-in-one control to text-to-image diffusion models.
\newblock In \emph{NeurIPS}, 2024.

\bibitem[Zhao et~al.(2023)Zhao, Bai, Zhu, Zhang, Xu, Zhang, Zhang, Meng, Timofte, and Van~Gool]{zhao2023ddfm}
Zixiang Zhao, Haowen Bai, Yuanzhi Zhu, Jiangshe Zhang, Shuang Xu, Yulun Zhang, Kai Zhang, Deyu Meng, Radu Timofte, and Luc Van~Gool.
\newblock Ddfm: denoising diffusion model for multi-modality image fusion.
\newblock In \emph{ICCV}, 2023.

\end{thebibliography}
}


\newpage

\twocolumn[
\centering
\Large
\textbf{\emph{TopoCellGen}: Generating Histopathology Cell Topology with a Diffusion Model} \\\vspace{0.05cm}{--- Supplementary Material ---}
\\
\vspace{1.5em}
]

\noindent In the supplementary material, we begin with notations for foreground and background in \cref{sec:fore_back_note}, followed by a description of the background knowledge about persistent homology in \cref{sec:background}. 
Next, we provide detailed introduction to our layout-guided pathology image generation part in \cref{sec:layout-to-image}, followed by the comprehensive descriptions of the datasets in \cref{sec:datasets}.
The implementation details are provided in \cref{sec:impl}. In \cref{sec:metrics}, we discuss the evaluation metrics in detail. To ensure the generation accuracy, we conduct the analysis on cell count distribution across training and test sets in \cref{sec:count_distribution_analysis}. More ablation studies are given in \cref{sec:addi_ablation_study}. The biological plausibility analysis by the domain expert is provided in \cref{sec:biological_plausibility}. Then, we provide the spatial point pattern analysis using multivariate Ripley's K-functions in \cref{sec:spatial_analysis}, followed by the discussion on computational cost and scalability of our method in \cref{sec:cc_scale}.
Finally, we discuss the limitations of our method in \cref{sec:limitations}.

\section{Notes on Foreground and Background}
\label{sec:fore_back_note}
Here, we provide some notations about foreground and background in our paper. In our experiments, the background of the layouts is black (the pixel value of 0) as can be seen in \cref{fig:merged-overview} and \cref{fig:qualitative_results}. 
For better visualization, we display the multi-class cell layouts with white as the background in 
\cref{fig:motivation}, \cref{fig:TopoFD_intuition_v2} and \cref{fig:TopoFD_pipeline} of the main paper.

\section{Background: Persistent Homology}
\label{sec:background}
In algebraic topology~\cite{munkres1984elements}, homology classes capture topological features across different dimensions. For instance, 0-, 1-, and 2-dimensional structures represent connected components, loops (or holes), and voids, respectively. In binary images, the number of $d$-dimensional topological features is described by the $d$-dimensional Betti number, $\beta_d$.\footnote{Technically, $\beta_d$ measures the dimension of the $d$-dimensional homology group. The number of distinct homology classes is exponential in $\beta_d$.} While topological structures are well-defined in binary images, extending this theory to real-world data, which is often continuous and noisy, poses challenges.

In the case of analyzing cell point clouds, where data is inherently discrete, we require a robust framework to infer the underlying topological structures. Persistent homology, developed in the early 2000s~\cite{edelsbrunner2002topological, edelsbrunner2022computational}, addresses this need by tracking the evolution of topological features across multiple scales.

Given a point cloud in the 2D space \mbox{$P \subseteq \mathbb{R}^2$}, a filtration is built by considering a growing family of simplicial complexes constructed from the point cloud as a function of a parameter (e.g., radius). For each parameter value, we define a set of simplices connecting the points, starting from isolated vertices and gradually adding edges and higher-dimensional simplices as the parameter increases. This creates a series of nested simplicial complexes: $ \varnothing \subseteq K_{r_1} \subseteq K_{r_2} \subseteq ... \subseteq K_{r_n}$. As the parameter grows, the topology of the complexes changes, with new connected components and loops emerging or vanishing.

Persistent homology captures these changes, tracking the birth and death of topological features over the filtration. The result is summarized in a persistence diagram (Dgm), which provides a multi-scale representation of the topological structures. A Dgm consists of points in a 2D plane, each representing a topological feature. The coordinates of each point, $(b, d)$, correspond to the feature's birth and death filtration values, providing a concise description of its persistence across scales.

\section{Layout To Image Generation}
\label{sec:layout-to-image}
\begin{figure}[htbp]
    \centering
    \includegraphics[width=0.48\textwidth]{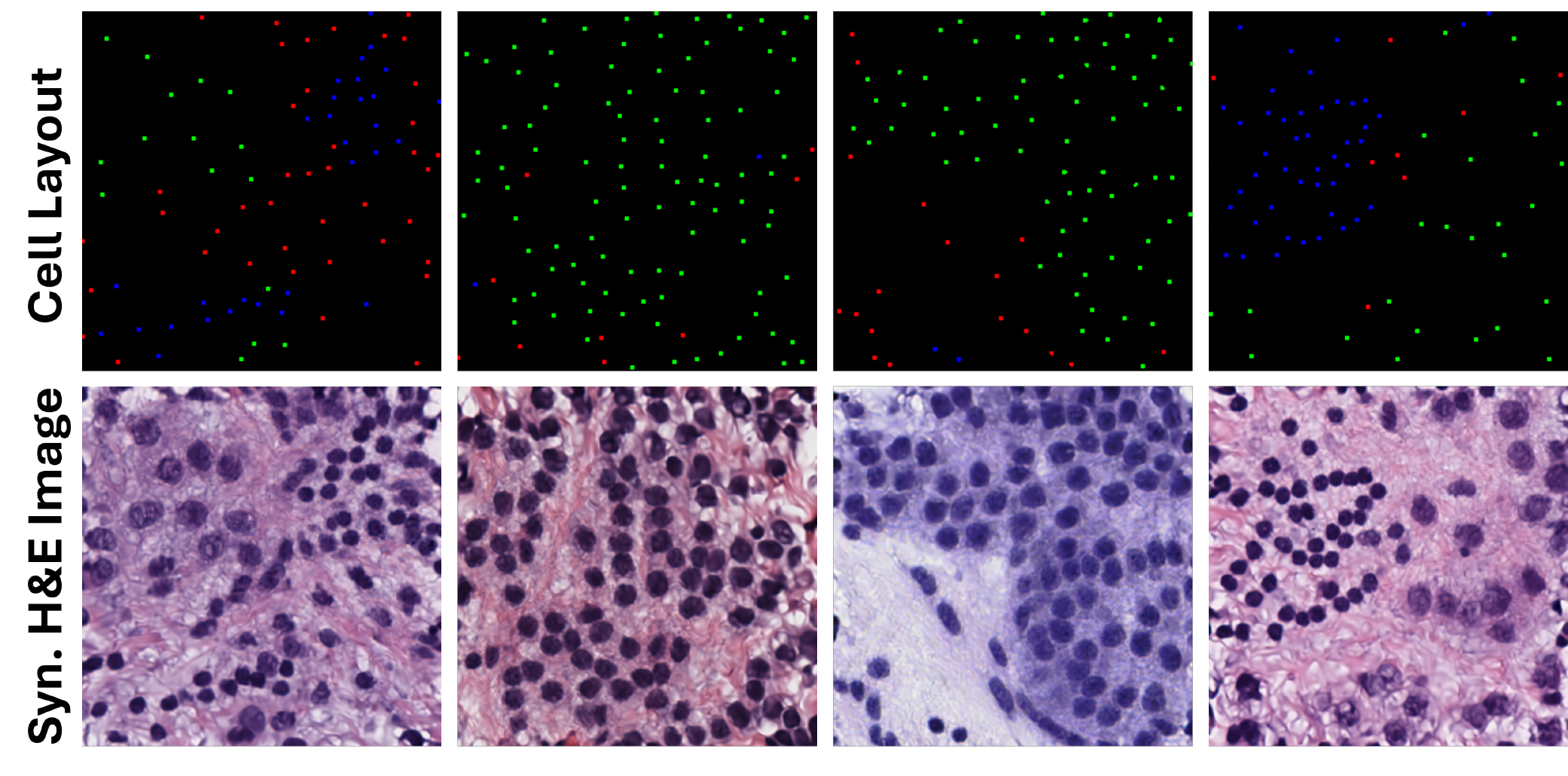}
    \caption{Qualitative results generated by our layout and image generation framework for downstream tasks.}
    \label{fig:layout_and_image}
\end{figure}

In this section, we introduce our layout-guided image generation framework in detail. The framework leverages a guided diffusion model (ADM) \cite{dhariwal2021diffusion} to generate H\&E images conditioned on multi-class cell layouts. The layouts serve as explicit conditional inputs to the diffusion model, which learns to reconstruct high-resolution pathology images from noisy counterparts during the reverse diffusion process. The conditioning mechanism is implemented using a cross-attention layer that seamlessly integrates cell layout information into the diffusion model. As shown in \cref{fig:layout_and_image}, the generated H\&E images generated by the model accurately depict the relative densities and arrangements of different cell types, while preserving the fine-grained details characteristic of histopathology images, such as nuclear shapes and staining patterns. This helps greatly improve the performance of downstream tasks, such as cell detection and classification.

\section{Details of the Datasets}
\label{sec:datasets}
\myparagraph{BRCA-M2C dataset} \cite{abousamra2021multi} is obtained from the TCGA dataset and contains $80$, $10$, and $30$ pathology image patches for training, validating, and testing, respectively. This dataset provides dot annotations for multi-class classification in breast cancer images. All images are around $500 \times 500$ pixels. The cell classes are lymphocytes, tumor or epithelial, and stromal cells.

\myparagraph{Lizard Dataset} \cite{graham2021lizard} is a large-scale resource for nuclear instance segmentation and classification, specifically targeting colonic tissue in computational pathology. It includes nearly $495,000$ manually and semi-automatically annotated nuclei, categorized into six classes: epithelial cells, connective tissue cells, lymphocytes, plasma cells, neutrophils, and eosinophils. $238$ images in the dataset are sourced from $6$ publicly available datasets, ensuring diverse representations of normal, inflammatory, dysplastic, and cancerous colonic conditions.

\section{Implementation Details}
\label{sec:impl}
Our work is mainly based on guided-diffusion (ADM)~\cite{dhariwal2021diffusion}. The condition of our model is a list of cell counts. An embedding of the condition is obtained by using an encoding network. After that, we feed this embedding to all the residual blocks in the network by adding it to the timestep embedding~\cite{nichol2021improved}. For every dataset, the image resolution is $256 \times 256$. Our diffusion models use a cosine noise scheduler~\cite{nichol2021improved}, with noising timesteps of $1000$ for training. We first pre-train the diffusion model using only $\mathcal{L}_{simple}$ for $150$K steps, then train with the three losses for $210$K steps. During the inference, we use $100$ steps of DDIM~\cite{song2020denoising}. The learning rate is $2\times 10^{-5}$ and the batch size is $5$. $\lambda_{c}$, $\lambda_{\text{intra}}$ and $\lambda_{\text{inter}}$ are all set to $0.0005$. 

For the layout-guided generation model, the learning rate is also $2\times 10^{-5}$ and we train the model only using $\mathcal{L}_{simple}$ for $360$K steps. The batch size is $6$. The image resolution is also $256\times256$. 
These experiments were conducted on 1 NVIDIA RTX A6000 GPU with 48GB RAM.

Our experiments designate specific test sets for each dataset to evaluate the synthetic cell layout generation process. For the BRCA-M2C dataset, we utilize $30$ images in the test set, which were pre-defined in the dataset. To prepare these for testing, each image is segmented into patches using a sliding window approach with a stride of $32$ pixels, resulting in patches of size $256\times256$. This process yields a total of $1,550$ patches for the BRCA-M2C dataset. We randomly select $20\%$ of the cell layouts as the test set for the Lizard dataset, which lacks predefined training and test splits. The chosen images undergo the same patching procedure, generating $256\times256$ patches, resulting in $1,000$ patches for the test set of the Lizard dataset.

In generating synthetic layouts, we aim to match the channel-wise cell counts observed in the real layouts of the test set. For each real test layout, we calculate the counts of each cell type across the channels and use these as conditional inputs during inference. This ensures that the generated synthetic patches exhibit similar cell count distributions to those observed in the real test layouts.

\section{Evaluation Metrics}
\label{sec:metrics}
To evaluate the quality of the generated cell layouts and pathology images, we employ a set of metrics focusing on different aspects, such as visual fidelity, topological similarity, and utility to downstream tasks.

First, the \textbf{Fréchet Inception Distance (FID)}~\cite{heusel2017gans} measures visual similarity by comparing the distributions of features extracted from a pre-trained Inception network between real and generated images. Lower FID scores indicate greater visual realism in the generated images. Feature extraction is tailored to each dataset with custom-trained models.
Here, the FID we used is the \textbf{spatial-FID} proposed in Spatial Diffusion \cite{li2024spatial}. The spatial-FID replaces visual features with a spatial representation derived from an autoencoder's intermediate layer, and we trained the autoencoder in the same way. In addition, we extended it to the Lizard dataset by training another autoencoder in the same manner.
We also evaluate the accuracy of the generation through cell count error, calculating discrepancies between real and generated cell counts per type and overall. In our experiments, we use the connected component labeling method~\cite{samet1988efficient} to count the cell numbers. Assume there 
are $n$ types of cells. For each cell type $i$, the \textbf{cell count error (CCE)} across $N$ test samples is defined as:
\begin{equation*}
    Cell\_Count\_Error^{(i)} = \frac{1}{N} \sum_{j=1}^{N} \left| c_{real,j}^{(i)} - c_{syn,j}^{(i)} \right|
\end{equation*}
with \textbf{total count error (TCE)} calculated as:
\begin{equation*}
    Total\_Count\_Error = \frac{1}{N} \sum_{j=1}^{N} \left| \sum_{i=1}^{n} c_{real,j}^{(i)} - \sum_{i=1}^{n} c_{syn,j}^{(i)} \right|
\end{equation*}
where $c$ is the cell count. In addition, our proposed \textbf{TopoFD} metric is used to evaluate the topological similarity between real and generated cell layouts. Lower TopoFD scores indicate closer alignment in spatial structure. 

We also use the metric proposed in \cite{wang2020topogan}, Maximum Mean Discrepancy (MMD) \cite{gretton2012kernel} to measure the topological difference between the real and synthetic distributions. The persistence diagrams from synthetic and real layouts are embedded into a reproducing kernel Hilbert space (RKHS). The MMD computes the distance between the mean embeddings of these two distributions in the RKHS. Given two sets of persistence diagrams, $\mathcal{D}_{syn}=\{{Dgm_i^{syn}}\}_{i=1}^{N}$ from the synthetic data and $\mathcal{D}_{real}=\{{Dgm_j^{real}}\}_{j=1}^{N}$ from the real data, we can define the mean of each diagram set, 
\begin{align*}
    \Phi(\mathcal{D}_{syn}):=\frac{1}{N}\sum_{i=1}^{N}\Phi(Dgm_i^{syn}) \\
    \Phi(\mathcal{D}_{real}):=\frac{1}{N}\sum_{j=1}^{N}\Phi(Dgm_i^{real})
\end{align*}
Then, the MMD is defined as:
\begin{equation*}
\text{MMD}(\mathcal{D}_{\text{syn}}, \mathcal{D}_{\text{real}}) := \|\Phi(\mathcal{D}_{\text{syn}}) - \Phi(\mathcal{D}_{\text{real}})\|_{\mathcal{H}} \
\end{equation*}
In terms of the kernel for persistence diagrams, we use the Gaussian kernel based on the 1-Wasserstein distance between diagrams, 
\begin{equation*}
k_{W_1}(Dgm_i, Dgm_j) = \exp\left(-\frac{W_1(Dgm_i, Dgm_j)}{\sigma^2}\right)
\end{equation*}

Lastly, to enhance downstream utility, we used $2,000$ generated image-layout pairs as augmented training data for cell detection and classification tasks, evaluating their performance with the \textbf{F1-score}.

\section{Cell Count Distribution Analysis}
\label{sec:count_distribution_analysis}
\begin{figure}[htbp]
    \centering
    \includegraphics[width=0.49\textwidth]{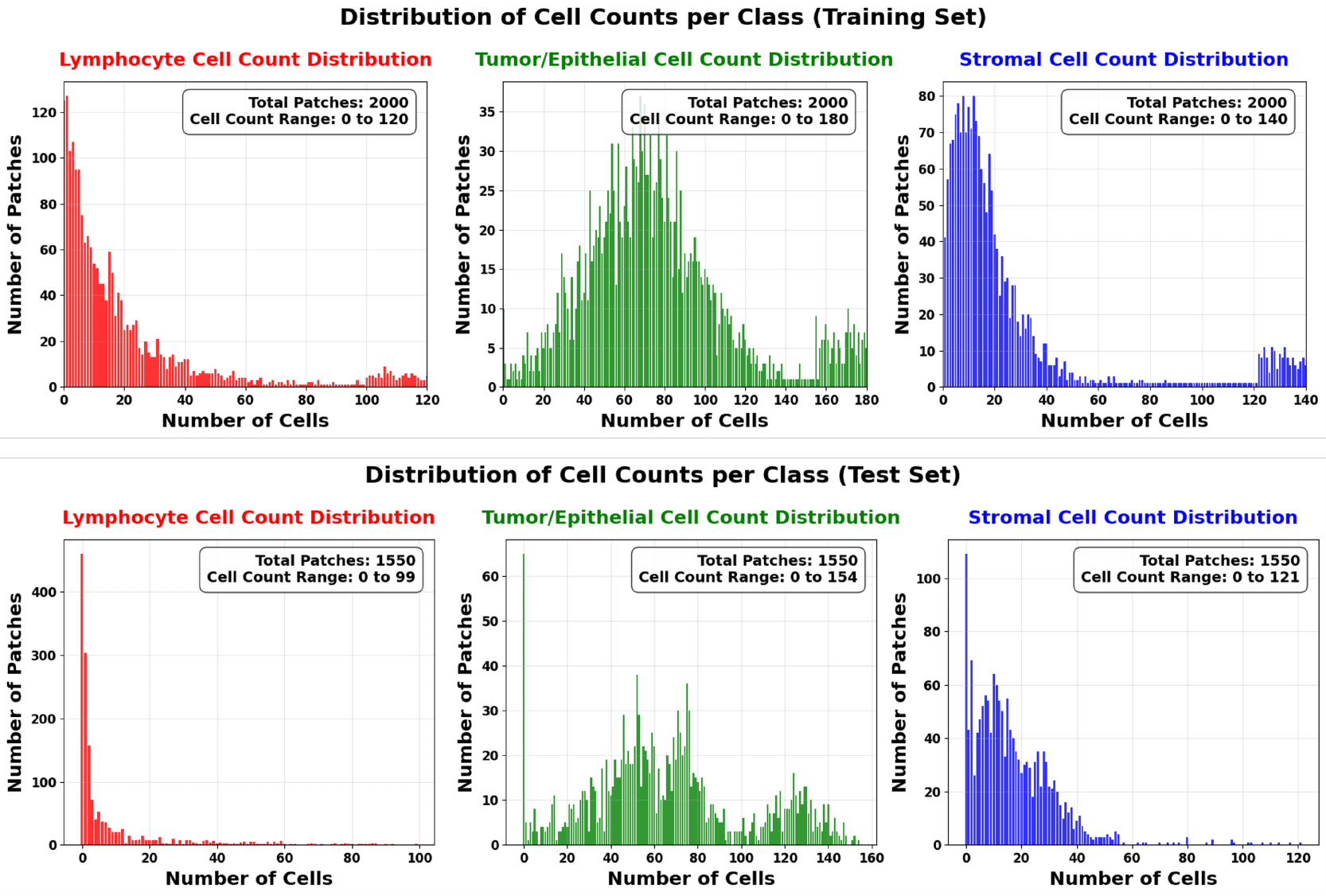}
    \caption{The statistical analysis of the cell count distributions on the BRCA-M2C training and test sets.}
    \label{fig:cell_count_statistics}
\end{figure}

Also, to ensure the accurate generation of cell distributions, the training set encompasses a wide range of cell count values. As shown in \cref{fig:cell_count_statistics}, we randomly select $2,000$ patches during the training. We analyzed and observed each cell type's range of cell counts in the training patches to confirm coverage across typical values observed in test conditions. This observation is crucial for the diffusion model, as it needs exposure to the range of cell counts during training to accurately generate corresponding counts during the inference.

\section{Additional Ablation Study}
\label{sec:addi_ablation_study}
\myparagraph{Ablation Study on learning rate.}
This ablation study examines the effect of different learning rates on model performance. The results indicate that a learning rate of $2\times10^{-5}$ achieves the best overall performance across all metrics, with the lowest FID, Total Counting Error, and TopoFD values. Higher learning rates, such as $1\times10^{-4}$, result in a higher total counting error and TopoFD, suggesting that an overly large learning rate may hinder convergence. Conversely, lower learning rates, including $1\times10^{-5}$ and $5\times10^{-5}$, show some improvements but do not reach the optimal balance across all metrics. The chosen learning rate of $2\times10^{-5}$, therefore, appears to provide the best trade-off, facilitating convergence that enhances both cell counting accuracy and fidelity in the synthetic cell layouts.
\begin{table}[ht]
\centering
\footnotesize
\begin{tabular}{cccc}
\hline
\multirow{2}{*}{learning rate} & \multicolumn{3}{c}{BRCA-M2C} \\ \cline{2-4} 
 & FID $\downarrow$ & TCE $\downarrow$& TopoFD $\downarrow$\\ \hline \hline
1e-4 & 0.021 & 12.357 & 75.667 \\
1e-5 & 0.015 & 6.314 & 81.397 \\
5e-5 & 0.066 & 12.367 & 85.949 \\
2e-5 & \textbf{0.005} & \textbf{5.192} & \textbf{69.354} \\ \hline
\end{tabular}
\caption{Ablation study on learning rate.
}
\label{ablation:lr}
\end{table}

\section{Biological Plausibility}
\label{sec:biological_plausibility}
\begin{figure}[h]
    \centering
    \vspace{-.05in}
    \includegraphics[width=0.49\textwidth]{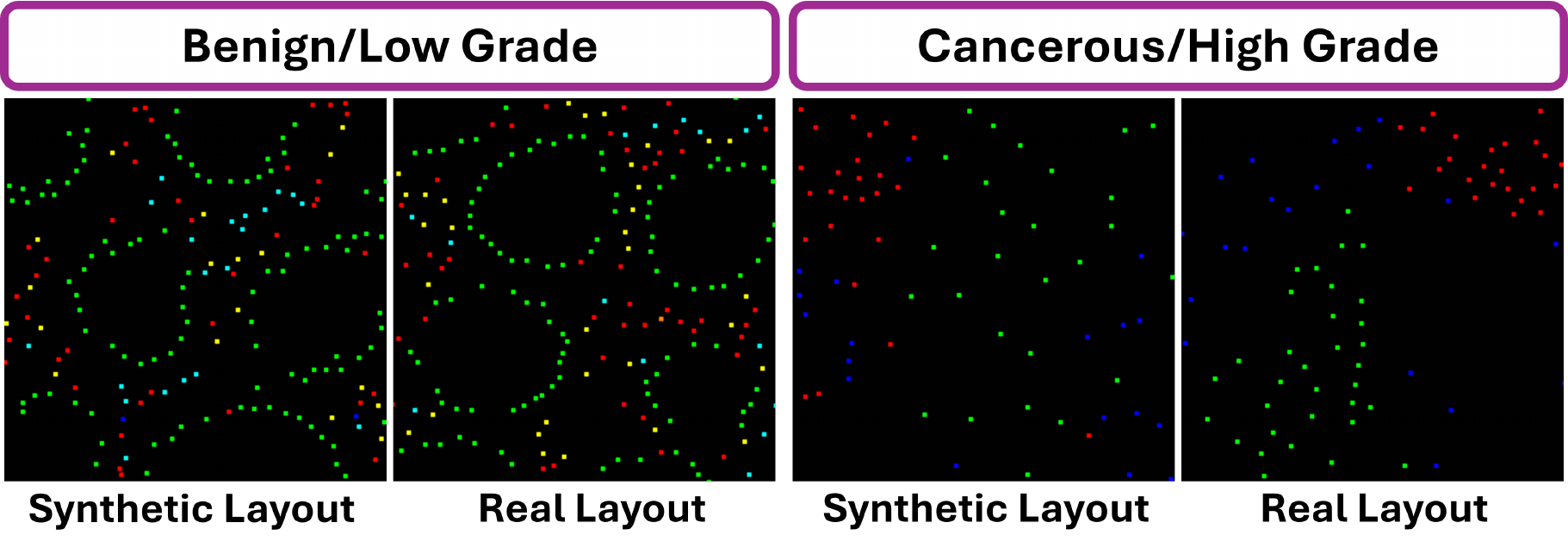}
    \vspace{-.3in}
    \caption{Biological plausibility validated by the domain expert.}
    \vspace{-.22in}
    \label{fig:biological_plausibility}
\end{figure}

Specifically, we randomly selected 10 pairs of real and synthetic cell layouts as shown in \cref{fig:biological_plausibility}. 
Without revealing their type (synthetic/real), we asked the expert to (1) identify which layout is synthetic; (2) characterize the tissue biology of these layouts.
The expert achieves only a $60\%$ accuracy in identifying the synthetic layout, confirming the realism of our synthetic layouts even to a domain expert. 
Regarding the characterization of tissue biology, as shown in \cref{fig:biological_plausibility}, the pathologist concluded that for each pair of layouts, the synthetic layout preserved the defining biological characteristics of its corresponding real sample, consistently reflecting benign/low-grade or cancerous/high-grade properties.
These experiments with a domain expert offer direct evidence, beyond quantitative measures and downstream analyses, that our generated layouts align well with actual biological structures.

\section{Spatial Point Pattern Analysis}
\label{sec:spatial_analysis}
We also evaluate our synthetic layouts using one standard statistical method for spatial point pattern analysis. Specifically, we employ multivariate Ripley's K-functions to evaluate the synthetic layouts of the BRCA-M2C dataset, which comprises $3$ cell types. For each test reference layout, we have a corresponding synthetic layout and extract cell centroids from both. We then compute $3$ K-functions to capture intra-class clustering (one per cell type) and $6$ cross-K functions to describe inter-class interactions.
Next, we examine the difference between real and synthetic K-values over $6$ radii: $[15, 30, 45, 60, 75, 90]$. For each radius and each cell-type pair, we perform a paired t-test to check if synthetic data deviates significantly from real layouts. This procedure yields $54$ p-values ($18$ from intra-class and $36$ from inter-class analyses). We then count the number of cases where these p-values exceed $0.05$, indicating no statistically significant difference.
Overall, as shown in \cref{tab:intra_class_spatial_analysis} and \cref{tab:inter_class_spatial_analysis}, \textbf{\textit{TopoCellGen}} achieves a greater number of radii with no significant difference is observed, compared to other methods. It most accurately produces both intra-class clustering and inter-class interactions, demonstrating the closest alignment with real data across the evaluated radii.

\begin{table}[h]
\centering
\tiny            
\setlength{\tabcolsep}{10pt}   
\renewcommand{\arraystretch}{0.9}  
\begin{tabular}{ccccc}
\hline
\multirow{2}{*}{Method} & \multicolumn{3}{c}{BRCA-M2C} &  \\ \cline{2-5} 
 & Lym. -- Lym. & Epi. -- Epi. & Stro. -- Stro. & Total \\ \hline \hline
ADM & 0/6 & 0/6 & 2/6 & 2/18 \\
TMCCG & 2/6 & 1/6 & 2/6 & 5/18 \\
Spatial Diffusion & 1/6 & 3/6 & 2/6 & 6/18 \\
\textit{TopoCellGen} & 3/6 & 5/6 & 4/6 & 12/18 \\ \hline
\end{tabular}
\vspace{-.1in}
\caption{Number of radii with no statistically significant difference ($p>0.05$) for intra-class spatial clustering.}
\vspace{-.2in}   
\label{tab:intra_class_spatial_analysis}
\end{table}

\begin{table}[h]
\centering
\tiny                           
\setlength{\tabcolsep}{2.5pt}     
\renewcommand{\arraystretch}{0.9}  
\begin{tabular}{cccccccc}
\hline
\multirow{2}{*}{Method} & \multicolumn{7}{c}{BRCA-M2C} \\ \cline{2-8} 
 & Lym. -- Epi. & Lym. -- Stro. & Epi. -- Lym. & Epi. -- Stro. & Stro. -- Lym. & Stro. -- Epi. & Total \\ \hline \hline
ADM & 1/6 & 3/6 & 2/6 & 1/6 & 1/6 & 2/6 & 10/36 \\
TMCCG & 3/6 & 2/6 & 3/6 & 4/6 & 3/6 & 2/6 & 17/36 \\
Spatial Diffusion & 3/6 & 4/6 & 2/6 & 3/6 & 1/6 & 2/6 & 15/36 \\
\textit{TopoCellGen} & 4/6 & 3/6 & 4/6 & 5/6 & 3/6 & 5/6 & 24/36 \\ \hline
\end{tabular}
\vspace{-.1in}
\caption{Number of radii with no statistically significant difference ($p>0.05$) for inter-class spatial interactions.}
\vspace{-.1in}    
\label{tab:inter_class_spatial_analysis}
\end{table}

\section{Computational Costs and Scalability}
\label{sec:cc_scale}
Currently, our model is trained on a single NVIDIA A6000 GPU with $48$ GB of memory for approximately $360$K steps, using a batch size of $5$ at $256\times256$ resolution within $200$ hours. The experiments can also be seamlessly scaled with data parallel training.

\section{Limitations}
\label{sec:limitations}
Our proposed \emph{TopoCellGen} will fail in some cases. First, the model is limited by its dependence on the cell types present in the training data, preventing it from generating layouts containing unseen cell types. Additionally, the model currently generates cell layouts in $256\times256$ patches, which constrains its application to small-scale regions.

\end{document}